\documentclass[acmsmall,screen,table,xcdraw]{acmart}

\usepackage{graphicx}
\usepackage{enumitem}
\usepackage{soul} 
\usepackage{xspace} 
\usepackage{subcaption} 
\usepackage{multirow} 
\usepackage{siunitx}
\usepackage[many]{tcolorbox} 
\usepackage{xurl}
\usepackage{listings} 

\AtBeginDocument{%
  \providecommand\BibTeX{{%
    \normalfont B\kern-0.5em{\scshape i\kern-0.25em b}\kern-0.8em\TeX}}}

\acmJournal{TOSEM}

\usepackage{amsmath}
\usepackage[linesnumbered,ruled]{algorithm2e}

\newcommand{\dapp}{\DH App\xspace}
\newcommand{\dapps}{\DH Apps\xspace}

\newtcolorbox{mybox}[2][]{
  top=0.15in,left=4pt,right=4pt,bottom=4pt,
  fonttitle=\bfseries,
  colbacktitle=gray,
  colback=gray!5,
  colframe=gray!40!black,
  enhanced,
  attach boxed title to top left={xshift=1.5em,yshift=-\tcboxedtitleheight/2},
  boxed title style={size=small},
  drop shadow={black!50!white},
  title=#2,#1
}

\colorlet{punct}{red!60!black}
\definecolor{background}{HTML}{EEEEEE}
\definecolor{delim}{RGB}{20,105,176}
\colorlet{numb}{magenta!60!black}

\lstdefinelanguage{json}{
    basicstyle=\scriptsize\ttfamily,
    numbers=left,
    numberstyle=\scriptsize,
    stepnumber=1,
    numbersep=8pt,
    showstringspaces=false,
    breaklines=true,
    frame=lines,
    backgroundcolor=\color{background},
    literate=
     *{0}{{{\color{numb}0}}}{1}
      {1}{{{\color{numb}1}}}{1}
      {2}{{{\color{numb}2}}}{1}
      {3}{{{\color{numb}3}}}{1}
      {4}{{{\color{numb}4}}}{1}
      {5}{{{\color{numb}5}}}{1}
      {6}{{{\color{numb}6}}}{1}
      {7}{{{\color{numb}7}}}{1}
      {8}{{{\color{numb}8}}}{1}
      {9}{{{\color{numb}9}}}{1}
      {:}{{{\color{punct}{:}}}}{1}
      {,}{{{\color{punct}{,}}}}{1}
      {\{}{{{\color{delim}{\{}}}}{1}
      {\}}{{{\color{delim}{\}}}}}{1}
      {[}{{{\color{delim}{[}}}}{1}
      {]}{{{\color{delim}{]}}}}{1},
}

\newcommand{\countobservations}{
    \def \countobservations{1}
}
\newcounter{observation}
\newenvironment{observation}[1]
{   \refstepcounter{observation}
    \noindent \textbf{Observation~\theobservation) \textit{#1}}
}

\countobservations

\newcommand{\countimplications}{
    \def \countimplications{1}
}
\newcounter{implication}
\newenvironment{implication}[1]
{
    \refstepcounter{implication}
    \noindent \textbf{Implication~\theimplication ) \textit{#1}}
}

\countimplications

\newcommand\RQOne{How long does it take to process a transaction in Ethereum?}
\newcommand\RQTwo{How accurate are the estimates for transaction processing time provided by Etherscan and EthGasStation?}

\graphicspath{{images/}}

\begin{document}

\title{Is my transaction done yet? An empirical study of transaction processing times in the Ethereum Blockchain Platform}


\author{Michael Pacheco}
\email{mpacheco@cs.queensu.ca}
\affiliation{%
  \institution{Software Analysis and Intelligence Lab (SAIL) at Queen's University}
  \city{Kingston}
  \country{Canada}
}

\author{Gustavo A. Oliva}
\email{gustavo@cs.queensu.ca}
\affiliation{%
  \institution{Software Analysis and Intelligence Lab (SAIL) at Queen's University}
  \city{Kingston}
  \country{Canada}
}

\author{Gopi Krishnan Rajbahadur}
\email{krishnan@cs.queensu.ca}
\affiliation{%
  \institution{Centre for Software Excellence at Huawei}
  \city{Kingston}
  \country{Canada}
}

\author{Ahmed E. Hassan}
\email{ahmed@cs.queensu.ca}
\affiliation{%
  \institution{Software Analysis and Intelligence Lab (SAIL) at Queen's University}
  \city{Kingston}
  \country{Canada}
}

\renewcommand{\shortauthors}{Pacheco, et al.}

\begin{abstract}
  Ethereum is one of the most popular platforms for the development of blockchain-powered applications. These applications are known as \dapps. When engineering \dapps, developers need to translate requests captured in the front-end of their application into one or more smart contract transactions. Developers need to pay for these transactions and, the more they pay (i.e., the higher the gas price), the faster the transaction is likely to be processed. Developing cost-effective \dapps is far from trivial, as developers need to optimize the balance between cost (transaction fees) and user experience (transaction processing times). Online services have been developed to provide transaction issuers (e.g., \dapp developers) with an estimate of how long transactions will take to be processed given a certain gas price. These estimation services are crucial in the Ethereum domain and several popular wallets such as Metamask rely on them. However, despite their key role, their accuracy has not been empirically investigated so far. In this paper, we quantify the transaction processing times in Ethereum, investigate the relationship between processing times and gas prices, and determine the accuracy of state-of-the-practice estimation services. Our results indicate that transactions are processed in a median of 57s and that 90\% of the transactions are processed within 8m. We also show that higher gas prices result in faster transaction processing times with diminishing returns. In particular, we observe no practical difference in processing time between expensive and very expensive transactions. With regards to the accuracy of processing time estimation services, we observe that they are equivalent. However, when stratifying transactions by gas prices, we observe that Etherscan's Gas Tracker is the most accurate estimation service for very cheap and cheap transaction. EthGasStation's Gas Price API, in turn, is the most accurate estimation service for regular, expensive, and very expensive transactions. In a post-hoc study, we design a simple linear regression model with only one feature that outperforms the Gas Tracker for very cheap and cheap transactions and that performs as accurately as the EthGasStation model for the remaining categories. Based on our findings, \dapp developers can make more informed decisions concerning the choice of the gas price of their application-issued transactions.
\end{abstract}

\ccsdesc[500]{General and reference~Empirical studies}
\ccsdesc[500]{General and reference~Estimation}
\ccsdesc[500]{Computer systems organization~Distributed architectures}

\keywords{Transaction Processing Time, Decentralized Applications (DApps), Ethereum, Blockchain}

\maketitle

\section{Introduction}
\label{sec:intro}

Blockchain is a novel software technology that enables secure and decentralized processing of digital transactions. The first mainstream blockchain platform was Bitcoin, which popularized the concept of cryptocurrencies. In the Bitcoin platform, the cryptocurrency is also called bitcoin (with a lowercase `b') and it is represented by the code BTC. The primary purpose of the Bitcoin platform is to enable the transfer of BTCs among user accounts. That is, the Bitcoin platform provides a platform for the processing of \textit{cryptocurrency transactions}. 

After Bitcoin, many other blockchain platforms have been developed. A special class of these platforms known as \textit{programmable blockchains} has recently gained particular notoriety. Different from Bitcoin, programmable blockchains also host and execute \textit{smart contracts} in addition to supporting cryptocurrency transactions. A smart contract is a stateful, general purpose computer program that is typically written with a high-level, object-oriented programming language (e.g., Solidity). One of the most popular programmable blockchain platforms is \textit{Ethereum}. In Ethereum, a user account can send \textit{contract transactions}. A contract transaction triggers the execution of a function defined in a smart contract.

Programmable blockchains enable the development of \textit{blockchain-powered applications}. In the world of Ethereum, these applications are known as \textit{decentralized applications} or simply \dapps. Due to the inherent properties of a blockchain (e.g., security, distributed processing), \dapps have the potential to transform how businesses currently operate. Indeed, this transformational potential yielded a critical demand for professionals with blockchain expertise. A recent report by LinkedIn \citep{anderson_2020} states: \textit{Last year, cloud computing, artificial intelligence, and analytical reasoning led LinkedIn's global list of the most in-demand hard skills. They're all on the list again this year, but a skill we weren't even looking at a year ago -- blockchain -- tops the list of most in-demand hard skills for 2020.} 

When engineering a \dapp, developers need to translate requests captured in the frontend of their application into one or more contract transactions. For example, assume that a finance company wishes to develop a bank \dapp on top of Ethereum. The developers of this bank application would thus need to \textit{translate} financial operations (e.g., pay a bill) into one or more contract transactions. In order to deliver a pleasant end-user experience, these transactions need to be processed as quickly as possible by the nodes that maintain the blockchain. Yet, the actual amount of time that it takes to process a transaction in Ethereum depends on several factors, including: the gas price set for the transaction (an Ethereum-specific form of transaction fees), the blockchain utilization level (i.e., how large the current workload is), and the transaction prioritization algorithms employed by the miner nodes (i.e., those entities that select and effectively process transactions in the blockchain). In other words, despite the critical role of transaction processing time in the final end-user experience, determining such time is far from trivial.

Out of the three aforementioned factors influencing transaction processing time, only the gas price can be controlled by the transaction issuer (e.g., \dapp developers). In the above described bank example, developers would likely achieve fast transaction processing times by setting a very high gas price. However, setting high gas prices for all transactions would likely render the application economically unviable. \textbf{In other words, the challenge is to dynamically determine the cheapest gas price that will provide the best possible end-user experience (transaction processing time).}

Online services have been developed to support transaction issuers (e.g., \dapp developers) in choosing appropriate gas prices. Currently, the two most popular services are Etherscan and EthGasStation. These services provide real time estimates of processing times for a given gas price (or set of gas prices). The rationale is that, by analyzing these estimates, transaction issuers can make a more informed gas price choice. Despite the popularity of the two aforementioned services, the accuracy of their processing time estimates remains unclear. In addition, Etherscan's service is proprietary and black box (i.e., its internal workings are undisclosed, preventing an interpretation of how the model operates).

In this study, we empirically investigate transaction processing times in Ethereum. More specifically, we determine the typical processing times, investigate the relationship between processing times and gas prices, and evaluate the accuracy of processing time estimation services. In the following, we list our research questions and the key results that we obtained:

\begin{itemize}[wide = 0pt, itemsep = 3pt, topsep = 3pt]
    \item \textbf{RQ1: \RQOne} \textit{Transactions are processed in a median of 57s. Also, 90\% of them are processed within 8m. We also observe that higher gas prices result in fast transaction processing times with diminishing returns (e.g., there is no practical difference between the processing times of expensive and very expensive transactions).}
    
    \item \textbf{RQ2: \RQTwo} \textit{Etherscan and EthGasStation use two prediction models each. Our results show that the four studied models are equivalent with a median absolute error in the range of 40.8s to 58.2s. However, in a stratified analysis based on gas price categories, we observe that the Etherscan Gas Tracker (proprietary, black box) is the most accurate model for very cheap and cheap transactions. The EthGasStation Gas Price API, in turn, is the most accurate model for the remaining price categories (regular, expensive, and very expensive.)}
\end{itemize}

Based on the results from RQ1 and RQ2, we conducted a post-hoc study in which we aimed at designing a simple and interpretable model that was at least as accurate as the existing top-performing models. In such a study, we show that a simple linear regression model that builds on only one feature is able to perform at least as accurately as the top-performing models for all price categories. In particular, our model outperforms the Etherscan Gas Tracker for \textit{very cheap} and \textit{cheap} transactions, which are the most difficult ones to  predict the processing time for.

The results of our paper support \dapp developers in making more informed decisions concerning the gas price of their application-issued transactions. Furthermore, our descriptive statistics of processing times in Ethereum should be of value to those who are considering the development of \dapps on top of this blockchain platform. 

The contributions of our study are as follows: (i) designing an approach to collect transaction processing times, which enables future studies in the area, (ii) characterizing transaction processing times for different gas price categories (\textit{very cheap}, \textit{cheap}, \textit{regular}, \textit{expensive}, and \textit{very expensive}), (iii) determining how accurate the existing processing time estimation services are, and (iv) developing a model that outperforms the existing estimation services. A supplementary package with the data analyzed in this study is made available online\footnote{\url{https://bit.ly/2YzfcKt}. For the final version of the paper, the data will be made available through a permanent link to a GitHub repository.}.

\medskip \noindent \textbf{Paper organization.} This paper is organized as follows. Section \ref{sec:background} introduces the key concepts that we use throughout this paper. Section \ref{sec:motiv-example} describes a motivating example, which clarifies how a practitioner can use a processing time estimation service in practice. Section \ref{sec:tx-processing-time} describes how we compute transaction processing times. Section \ref{sec:data-collection} outlines the data collection process of our study. Section \ref{sec:results} presents the motivation, approach, and our findings for each research question. Section \ref{sec:post-hoc-study} presents our post-hoc study. Section \ref{sec:implications} discusses the implications of our findings. Section \ref{sec:related-work} presents related work. Section \ref{sec:threats} discusses the threats to the validity of our findings. Finally, Section \ref{sec:conclusion} concludes the study.
\section{Background}
\label{sec:background}

In this section, we introduce the key concepts that are used throughout our paper.

\subsection{Blockchain}
\label{subsec:blockchain}

A blockchain is essentially a ledger of all transactional activity occurring on the network. The ledger is stored on the network using blocks: an object which stores a unique set of transactions, all of which are identifiable by their own unique ID. The information within blocks cannot be changed unless all of the blocks that came after it are also changed. As a result, it is common for information in a block to become immutable after a certain amount of blocks have been appended after it.


\subsection{Transactions}
\label{subsec:transactions}

Transactions are the means through which users interact with a blockchain. In Ethereum, there are two types of transactions: \textit{user transactions} and \textit{contract transactions}. User transactions enable users to send cryptocurrency to one another. The cryptocurrency used in Ethereum is known as Ether (ETH). A contract transaction, in turn, enables one to execute a function defined in a smart contract deployed in Ethereum.

As in most blockchain platforms, transactions in Ethereum must be paid for. The transaction fee is given by \textit{gas usage} $\times$ \textit{gas price}. The gas usage corresponds to the amount of computing power that was needed in order to process a given transaction. While this computing power is fixed for user transactions (21 GWEI = 2.1E-8 ETH), it varies considerably for contract transactions. More specifically, each instruction in the bytecode of a smart contract burns a certain amount of gas units \citep{wood2019ethereum}. The gas price, in turn, is a parameter set by the transaction issuer. The gas price corresponds to the amount of Ether that one is willing to pay for each unit of burnt gas. In practice, the gas price parameter is a way for users to create an incentive for miners to process transactions at a higher priority \citep{signer2018gas}, as miners are rewarded for their processing efforts with the cost of the transaction fee. 

\smallskip \noindent \textbf{Gas and transaction fees.} Transactions sent within the Ethereum blockchain require a transaction fee to be paid before being sent. This fee is calculated by \textit{gas usage} $\times$ \textit{gas price}. The \textit{gas usage} term refers to the amount of computing power consumed to process a transaction. This value is constant for user transactions, costing 21 GWEI, equivalent to 2.1E-8 ETH. For contract transactions, this value is calculated depending on the bytecode executed within a smart contract, as each instruction has a specific amount of gas units required for it to execute \citep{wood2019ethereum}. The \textit{gas price} term defines the amount of Ether the sender is prepared to pay per unit of gas consumed to process their transaction. This parameter is set by the sender, and allows users to incentivize the processing of their transaction at a higher priority \citep{signer2018gas}, as miners which process the transaction are rewarded based on the transaction fee value. 

\begin{figure}
 \centering
     \includegraphics[width=.5\textwidth,keepaspectratio=true]{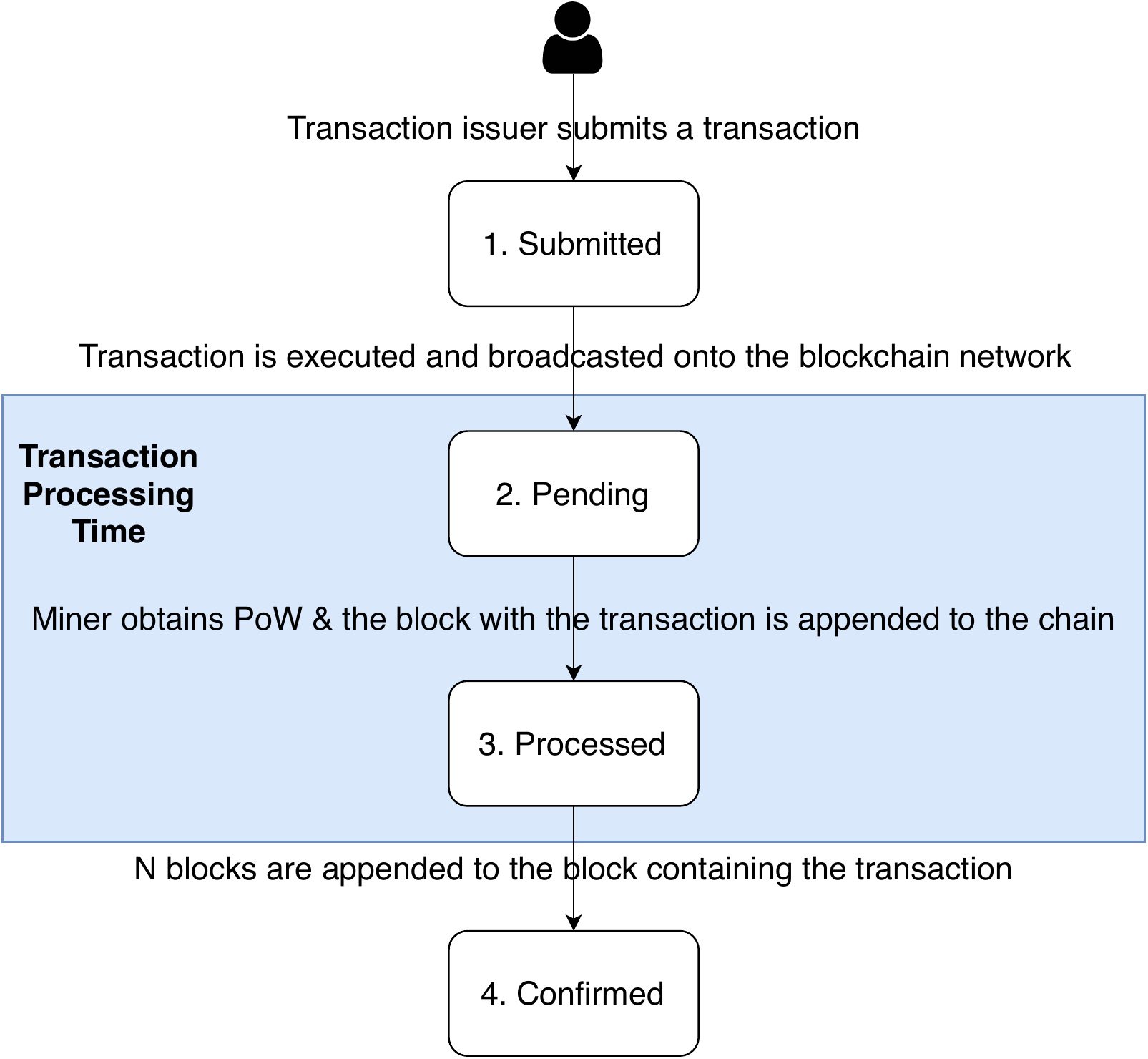}
     \caption{Transaction lifecycle in the Ethereum blockchain.}
     \label{fig:txn_lifecycle}
 \end{figure}

\medskip \noindent \textbf{Transaction processing lifecycle.} As illustrated in Figure \ref{fig:txn_lifecycle}, the lifecycle of a transaction $t$ begins when a transaction issuer submits it. Transaction $t$ is then broadcasted through the blockchain by means of the peer to peer network. Eventually, $t$ is discovered by most mining nodes in the network (each mining node has its own pool of pending transactions, or \textit{pending pool} for short). We consider $t$ to reach a \textit{pending} state when its existence is acknowledged by any miner node in the network. Eventually, a miner wins the Proof-of-Work \citep{Jakobsson99} competition\footnote{Proof-of-Work is a consensus protocol that requires nodes to solve a hard mathematical puzzle. The PoW consensus protocol ensures that there is no better strategy to find the solution to the mathematical puzzle than enumerating the possibilities (i.e., brute force). On the other hand, the verification of a solution is trivial and cheap. Ultimately, the PoW consensus protocol ensures that a trustworthy third-party (e.g., a bank) is not needed in order to validate transactions, enabling entities who do not know or trust each other to build a dependable transaction ledger.}, allowing such a miner to append a new block $b$ of transactions to the blockchain. These transactions are chosen from the miner's pending pool. If $t$ is included in $b$, we then consider $t$ to reach a \textit{processed} state as soon as the $b$ is appended to the blockchain (i.e., at the \textit{block timestamp} of $b$). We highlight that $n$ blocks need to be appended after $b$ in order for $t$ to be considered \textit{confirmed} (a.k.a., \textit{final}). There is no consensus on what the exact value of $n$ should be. The Ethereum whitepaper suggests $n = 7$ \citep{buterin2014ethereum}, which translates to approximately 01m 45s (since blocks are appended every 15s in average). Practitioners (end-users and developers) commonly use $n = 12$ as a rule of thumb\footnote{\url{https://stackoverflow.com/questions/49065176/how-many-confirmations-should-i-have-on-ethereum}}. Some crypto exchanges use higher thresholds, since they manipulate high-value crypto transactions and thus prefer to be conservative regarding confirmation. CoinCentral, for instance, uses $n = 250$, which translates to approximately 01h 02min 30s \citep{coincentral}.

\medskip \noindent \textbf{User-specific transaction ordering (nonce).} The \textit{nonce} is a transaction parameter that records the number of transactions that were previously sent by the transaction issuer (i.e., every time someone sends a new transaction, their nonce is increased by 1). This parameter exists to preserve transaction ordering in Ethereum. Hence, if a transaction issuer Charlie submits a very cheap transaction $t_1$ and, after 10 seconds, submits a very expensive transaction $t_2$, then $t_2$ can only be mined (i.e., added to a block) once $t_1$ has been mined. Since $t_1$ was submitted with a very cheap gas price, it is possible that $t_2$ will take long to be mined. In summary, a transaction with nonce $i$ can only be mined if the transaction with nonce $i-1$ has already been mined.
\section{Motivating Example}
\label{sec:motiv-example}

Assume that a DApp development organization (e.g., a company) wants to improve the end-user experience of their DApps. More concretely, assume that the organization wants to provide the following quality of service (QoS): 90\% of all their submitted transactions should be processed within 5 minutes. Such a quality of service would be somewhat easily achievable if the organization’s budget were infinite. Clearly, that is not the case for any real-world DApp development organization. Hence, let us assume that such an organization wishes to provide the aforementioned QoS while \textit{simultaneously} ensuring that the gas prices chosen for those transactions are as low as possible (i.e., minimize expenses). We refer to this balance as the \textit{time-expense balance}.

\smallskip \noindent \textbf{Using an estimation service in practice.} Meeting the organization's goal requires an accurate processing time estimation service. If a perfect estimation service existed, the company would be able to always achieve their desired time-expense balance. Let us show how this would work in practice. Assume that a developer named Charlie is responsible for programming the transaction submissions of a given DApp. Right before submitting any transaction, Charlie can request a \textit{lookup table} to the estimation service at timestamp \textit{t}, which would contain the processing time predictions for each possible value of gas price, based upon recently processed transactions in previous blocks before \textit{t}. Table~\ref{tab:lookup-table} shows an example of a hypothetical lookup table requested from an estimation service at timestamp \textit{t}.
For simplicity, we assume that the prices returned by the estimation service go from 1 to 60 GWEI in increments of 1. Because the purpose of the lookup table is to help Charlie choose a gas price for his transaction, we also assume that the estimated time of a transaction with a gas price $x_i$ is always higher than the estimated time of another transaction with price $x_{i+1}$, for the same timestamp \textit{t}. Finally we discretize prices into gas price categories to give a more practical, straightforward interpretation to the prices. We consider that [1,12] is very cheap, [13,24] is cheap, [25,36] is regular, [37,48] is expensive and [49,60] is very expensive. This allows Charlie to choose between prices to assign his transaction based on how quickly he wants it to be processed at that given time.

\begin{table}
    \centering
    \caption{A hypothetical lookup table of a processing time estimation service at a given timestamp \textit{t}.
    }
    \label{tab:lookup-table}
    \begin{tabular}{ccc}
    \hline
    \textbf{\begin{tabular}[c]{@{}c@{}}Gas Price \\ (GWEI)\end{tabular}} &
      \textbf{\begin{tabular}[c]{@{}c@{}}Gas Price \\ Category\end{tabular}} &
      \textbf{\begin{tabular}[c]{@{}c@{}}Estimated processing \\ time (minutes)\end{tabular}} \\ \hline
    1  & Very Cheap     & 12.0 \\
    2  & Very Cheap     & 9.0  \\
    3  & Very Cheap     & 7.0  \\
    4  & Very Cheap     & 6.5  \\
    5  & Very Cheap     & 6.0  \\
    6  & Very Cheap     & 5.0  \\
    7  & Very Cheap     & 4.5  \\
    8  & Very Cheap     & 4.2  \\
    9  & Very Cheap     & 4.0  \\
    10 & Very Cheap     & 3.9  \\
    …  &                & …    \\
    13 & Cheap          & 3.6  \\
    …  &                & …    \\
    25 & Regular        & 1.8  \\
    …  &                & …    \\
    37 & Expensive      & 1.0  \\
    …  &                & …    \\
    49 & Very Expensive & 0.7  \\
    …  &                & …    \\
    60 & Very Expensive & 0.5  \\ \hline
    \end{tabular}
    \end{table}

\smallskip \noindent \textbf{Evaluating the time-expense balance.} If a perfect estimation service existed, Charlie would achieve the desired time-expense balance by always picking the smallest price whose estimated processing time is shorter than or equal to 5 minutes. In Table~\ref{tab:lookup-table}, this value would be 6 GWEI. However, no such perfect estimation service exists in practice. In fact, devising an accurate estimation service is a major research challenge. Hence, let us assume that Charlie currently uses an estimation service $S$ that \textit{is not perfect} and that underestimates and overestimates with the same likelihood. Additionally, let us assume that Charlie has a \textit{fixed monthly budget} for transaction fees. Transaction fees depend on both gas usage and gas price of transactions. In the interest of simplification, let us assume that the transactions that Charlie submits always burn the same amount of gas units, such that the budget can be specified in terms of gas price only. Let us assume Charlie's budget is exactly 15,000 GWEI. Finally, to determine how good the estimation service $S$ is (i.e., how good the time-expense balance it provides is), Charlie computes the harmonic mean between the following two variables: (i) the percentage of submitted transactions that were processed within 5 minutes and (ii) how much of the budget is still left (as a percentage). The rationale behind using the harmonic mean is analogous to that of measuring the balance between precision and recall through the F1-measure (harmonic mean between precision and recall, which are both rates).

\begin{figure}
    \centering
    \includegraphics[width=0.8\linewidth]{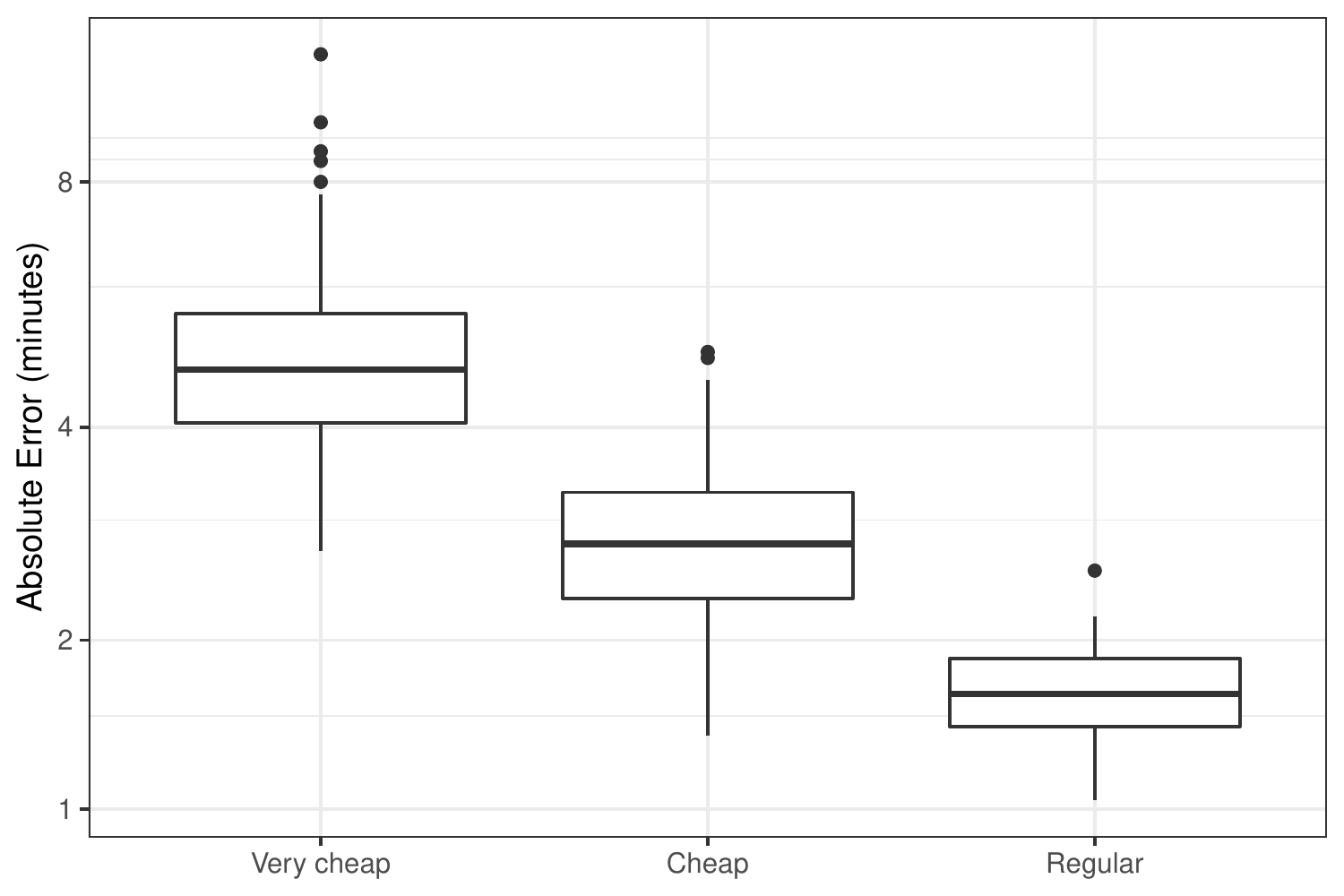}
    \caption{Absolute errors given by a hypothetical processing time estimation service (y-axis in log1p scale)}
    \label{fig:motiv-example-abs-error}
\end{figure}

Now that the context has been specified, let us simulate how Charlie would use $S$ for a month. Since $S$ is not perfect, Charlie does not trust it very much and thus always picks the \textit{third} smallest price that satisfies the processing time criterion (in Table~\ref{tab:lookup-table}, this would be 8 GWEI). For each transaction that Charlie submitted during the month, Charlie recorded both the predicted and the actual processing times. After submitting all transactions during that month, Charlie computed the absolute errors of the predictions (i.e., the absolute value of \textit{predicted} minus \textit{actual}). Figure~\ref{fig:motiv-example-abs-error}\footnote{The results shown in Figure~\ref{fig:motiv-example-abs-error} were produced using synthetic data. The script used for generating these data is available as part of our supplementary material.} summarizes the results that Charlie obtained.

As the figure indicates, the median error per category decreases from left to right. Also, the difference in the median absolute error between \textit{very cheap} and \textit{cheap} is larger than that between \textit{cheap} and \textit{regular}. In fact, as we shall see in RQ2, state-of-the-practice estimation services exhibit these two characteristics. With estimation service S, 7.8\% of the transactions were processed within 5 minutes and 47.8\% of the budget was still free. These results yield a time-expense balance of 13.4\%.

Now, let us assume that we could go back in time and make the estimation service fifty percent more accurate (i.e., absolute errors cut in half). We refer to this improved estimation service as $S^+$. What would be the difference in the time-expense balance be?  With the new improved estimation service, 46.6\% of the transactions would have been processed within 5 minutes and 27.1\% of the budget would have been left unused. The time-expense balance in this case would have been 34.2\% (2.55 times better than before). The results are summarized in Table~\ref{tab:motiv-example-summary}.

\begin{table}
    \centering
    \caption{Summary of the Time-Expense balance quality yielded by estimation services $S$ and $S^+$.}
    \label{tab:motiv-example-summary}
    \resizebox{\textwidth}{!}{%
    \begin{tabular}{clrrrrr}
    \hline
    \textbf{\begin{tabular}[c]{@{}c@{}}Estimation \\ Service\end{tabular}} &
      \multicolumn{1}{c}{\textbf{Choice}} &
      \multicolumn{1}{c}{\textbf{Budget}} &
      \multicolumn{1}{c}{\textbf{\begin{tabular}[c]{@{}c@{}}Total Spent \\ (GWEI)\end{tabular}}} &
      \multicolumn{1}{c}{\textbf{\begin{tabular}[c]{@{}c@{}}Budget \\ Free (\%)\end{tabular}}} &
      \multicolumn{1}{c}{\textbf{\begin{tabular}[c]{@{}c@{}}QoS: Ratio of Txns.\\ Processed Within \\ 5 minutes (\%)\end{tabular}}} &
      \multicolumn{1}{c}{\textbf{\begin{tabular}[c]{@{}c@{}}Time-Expense \\ Balance Quality (\%)\end{tabular}}} \\ \hline
    $S$ &
      \begin{tabular}[c]{@{}l@{}}3rd smallest \\ price\end{tabular} &
      15,000 &
      7,837 &
      47.8 &
      7.8 &
      13.4 \\
    $S^+$ &
      \begin{tabular}[c]{@{}l@{}}3rd smallest \\ price\end{tabular} &
      15,000 &
      10,940 &
      27.1 &
      46.6 &
      34.2 \\ \hline
    \end{tabular}%
    }
    \end{table}

\smallskip \noindent \textbf{Take-away.} With this motivating example, we show how a practitioner can use a processing time estimation service in practice. Most importantly, we also show that improvements in the accuracy of the estimations given by these services lead to a better time-expense balance, ultimately benefiting the end-users of DApps. 

We also clarify that, while the need to \textit{reduce} transaction processing times is likely higher for mission-critical DApp, any DApp that issues a reasonable amount of daily transactions would benefit from a better time-expense balance. Assume, for instance, that a given DApp does not have a strong QoS requirement. For example, assume that the QoS in question is having 90\% of the transactions processed within 30 minutes. Additionally, assume that developers commonly spend an average of 4 GWEI per transaction to achieve this QoS. If developers can achieve this same QoS while spending an average of 2 GWEI instead, the costs to run the DApp would be reduced by half, making it more cost-effective. The key challenge for DApp developers is thus spending as little as possible to achieve the desired QoS (i.e., optimizing the time-expense balance). To address this challenge, accurate processing time estimation services are needed.

In this paper, we aim to determine how long transactions normally take to be processed in Ethereum (RQ1) and how accurate the current existing services are (RQ2).
\section{Computing Transaction Processing Times}
\label{sec:tx-processing-time}

In this paper, we define the processing time of a transaction $t$ as the time elapsed from when a $t$ enters the \textit{pending state} until $t$ enters the \textit{processed state} (Figure \ref{fig:txn_lifecycle}). To compute the processing time of a transaction $t$, we thus need to determine two timestamps: the timestamp at which $t$ enters the \textit{pending} state (henceforth \textit{pending timestamp}) and the timestamp at which $t$ enters the \textit{processed} state (henceforth \textit{processed timestamp}). The processing time is then calculated by simply taking the delta between these two timestamps. 

We compute transaction processing times by mining the \textit{pending timestamp} and the \textit{processed timestamp} from the Etherscan website. In particular, we highlight that the transaction timestamp, as recorded in the blockchain, is an imprecise representation of the \textit{processed timestamp} and we thus refrain from using it. A detailed explanation of how we obtained the two aforementioned timestamps is described in Appendix \ref{appendix:tx-processing-time}. All processing times reported in this paper are given in \textit{minutes}.

\section{Data Collection}
\label{sec:data-collection}

In this section, we describe the data sources that we used (Section \ref{subsec:data-sources}) and the data collection steps that we followed (Section \ref{subsec:data-steps}) in order to answer our research questions.

\subsection{Data Sources}
\label{subsec:data-sources}

Our study involves three main data sources: Etherscan, EthGasStation, and Google BigQuery. In the following, we describe these data sources:

\medskip \noindent \textbf{Etherscan.} Etherscan is a widely used real-time dashboard for the Ethereum blockchain. This website contains information about transactions, blocks, and even the source code for many smart contracts. Most importantly for this paper, Etherscan also tracks pending transactions and provides two transaction processing time predictors: one in the Pending Transactions webpage and another in the Gas Tracker webpage.


\medskip \noindent \textbf{EthGasStation.} EthGasStation is an online service that provides estimates of transaction processing times for a range of gas prices. The predictions can be obtained from two different endpoints of their web service, which we refer to as the Gas Price API and the Prediction Table API.

\medskip \noindent \textbf{Google BigQuery.} Google BigQuery is an online platform that is used to analyze large datasets. Google also actively maintains several public datasets, including an Ethereum dataset\footnote{The dataset is called \texttt{bigquery-public-data.crypto\_ethereum}}. This dataset is updated daily and it contains metadata about transactions, blocks, and smart contracts (among others).

\subsection{Approach} 
\label{subsec:data-steps}

\begin{figure}
\centering
    \includegraphics[width=1.0\linewidth]{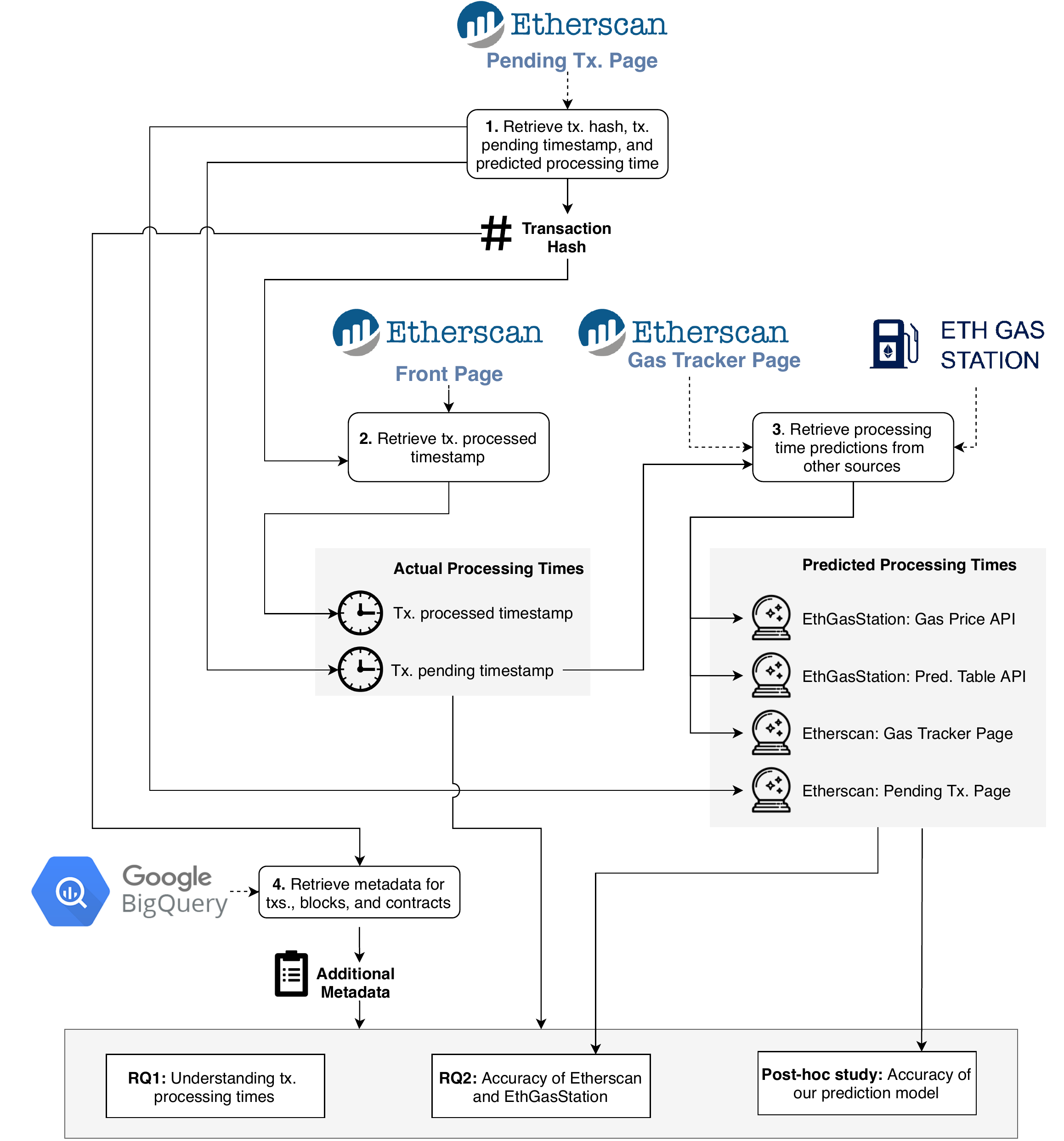}
    \caption{An overview of our data collection approach. Dashed lines indicate a connection to a data source.}
    \label{fig:data-collection}
\end{figure}

In the following, we discuss the specific pieces of data that we collected to help answer each of our research questions. An overview of our data collection approach is illustrated in Figure \ref{fig:data-collection}.

\smallskip \noindent \textbf{1. Retrieve tx. hash, tx. pending timestamp, and predicted processing time.} Every transaction has a unique identifier known as the \textit{transaction hash}. We collected this hash for all transactions shown in the Pending Transactions webpage of Etherscan during the period of Nov. 21st 2019 until Dec. 09th 2019. In total, we collected 283,102 hashes. These collected hashes represent all the transactions that we studied in this paper. They are used as input to all other data collection steps.

For each transaction in the Pending Transactions webpage, we also collect two pieces of information: the \textit{transaction pending timestamp} (see Section \ref{appendix:pending-timestamp}) and the predicted processing time (see the \textit{estimated confirmation duration} in Figure \ref{fig:appendix-tx-firstseen-lastseen}). Although Etherscan calls it Estimated *Confirmation* Duration, the value actually refers to the estimated processing time (i.e., the shown value does not include any extra time for transaction confirmation/finality). Indeed, hovering the mouse over the question mark symbol to the left of the Estimated Confirmation Duration text reveals a popup with the following message: ``An estimate of the duration before the transaction is mined. Time can vary according to the network congestion''.

\medskip \noindent \textbf{2. Retrieve tx. processed timestamp.} We retrieve the \textit{processed timestamp} for each studied transaction (see Section \ref{appendix:processed-timestamp}). At this point, we are able to compute the actual transaction processing times (i.e., the delta between the pending timestamp and the processed timestamp). The actual processing times are used in our two research questions.

\medskip \noindent \textbf{3. Retrieve processing time predictions from other sources.} We obtain three additional processing time predictions, which are provided by EthGasStation and Etherscan. We obtain predictions continuously, every 1 minute. We also save the \textit{retrieval timestamp}, which denotes the instant at which the predictions were retrieved. This retrieval timestamp allows us to relate the predictions to our studied transactions. This is achieved by mapping the \textit{pending timestamp} of our studied transactions to the closest \textit{retrieval timestamp} (with \textit{retrieval timestamp} $<$ \textit{pending timestamp}). We highlight that these three additional models only provide predictions for a range of specific gas prices at any given time. If there is no exact match between the gas price of a studied transaction and the set of specific gas prices listed by a model, we match the former to the closest price in the latter. In the following, we describe how we obtain the predictions from EthGasStation and Etherscan.

From EthGasStation, we collect predictions from its two API end points: Gas Price\footnote{\url{https://ethgasstation.info/json/ethgasAPI.json}} and Prediction Table\footnote{\url{https://ethgasstation.info/json/predictTable.json}}. The information provided by these endpoints is also found on the front page\footnote{\url{https://ethgasstation.info/}} and predict table\footnote{\url{https://ethgasstation.info/predictionTable.php}} webpage of EthGasStation respectively. Both of these predictions also limit the users to predictions based on a certain range of gas prices associated with transactions. In RQ2, we evaluate the predictions made for a specific gas price by these two endpoints and determine whether they provide the same results. These two EthGasStation models are \textit{open-source}, and can be found on multiple Github repositories\footnote{\url{https://github.com/ethgasstation/ethgasstation-backend}}\footnote{\url{https://github.com/ethgasstation/gasstation-express-oracle}}. 
EthGasStation also includes a Transaction Calculator (Tx Calculator)\footnote{\url{https://legacy.ethgasstation.info/calculatorTxV.php}} on their website, which suggests four different gas prices (categorized into \textit{cheap}, \textit{average}, \textit{fast}, and \textit{fastest}) and predicts the mean processing times for each of them. The four gas prices suggested by the Tx Calculator are identical to those suggested in the Gas Price API Endpoint at the same time of access, as they use the same Poisson model that is fit on transaction data every 100 blocks\footnote{\url{https://legacy.ethgasstation.info/FAQcalc.php}}. However, the corresponding processing time predictions of these two tools are actually different, as the Tx Calculator provides the mean processing time, while the Gas Price API Endpoint provides the exact value resulting from their Poisson model. The Gas Price Endpoint in turn is a more relevant source for comparing the predictions made by our model and EthGasStation’s Poisson model.

Etherscan contains a webpage called Gas Tracker\footnote{\url{https://etherscan.io/gasTracker}}. In this webpage, Etherscan provides a range of gas prices and a prediction of their resulting processing times. The Gas Tracker predictions give users a more general sense of potential processing times, as it limits users to a set of predefined gas prices to analyze. In contrast, the prediction made in the Pending Transactions webpage (step 1) is specific to a given transaction. We highlight that the prediction models underlying the Gas Tracker and the Pending Transactions webpages are proprietary, \textit{black boxes}. For instance, there is no public online documentation regarding which machine learning algorithm is being used, the features employed by the model, and the model retraining frequency.


The predicted processing times from the 4 different sources are compared to each other in RQ2.

\smallskip \noindent \textbf{4. Retrieve metadata for txs., blocks, and contracts.} We use Google BigQuery to obtain additional metadata from transactions and their associated blocks and contracts. These metadata allow us to obtain information associated with the collected transactions, such as gas price, block number, and network congestion. Metadata are obtained by querying the \textit{transactions}, \textit{blocks}, and \textit{contracts} tables. These metadata support answering RQ1 and RQ2 (e.g., by observing how gas price relates to actual and predicted processing times). We also leverage these metadata to engineer the features of the model that we propose in the post-hoc study.

\smallskip
\begin{mybox}{Summary}
    \begin{itemize}[itemsep = 3pt, label=\textbullet, wide = 0pt]
	    \item Analysis period: November 21, 2019 until December 09, 2019 (18 days).
		\item Data sources: Etherscan, EthGasStation, and Google BigQuery.
        \item Pieces of data collected: Transaction processing times, predicted transaction processing times from Etherscan (Pending Tx. and Gas Tracker webpages) and EthGasStation (Gas Price and Prediction Table APIs). Metadata for transactions, blocks, and smart contracts.
    \end{itemize}
\end{mybox}
\section{Results}
\label{sec:results}

\subsection{RQ1: \RQOne}
\label{subsec:rq1}

\noindent \textbf{Motivation.} In Ethereum, users (e.g., developers) interact with the blockchain via transactions. Developers of \dapps want the transactions issued by their applications to be processed as quickly as possible in order to deliver a pleasant end-user experience. Yet, \dapp developers need to pay for these transactions by assigning a certain gas price to them. That is, \dapp developers need to find the optimal balance between cost (transaction fees) and end-user experience (transaction processing times). Nonetheless, to this day, there is little empirical evidence to guide developers in making informed decisions regarding the gas price of transactions. Even simple statistics such as the typical processing time of transactions have been little investigated.



\smallskip \noindent \textbf{Approach.} As described in Section \ref{sec:data-collection} and summarized in Figure \ref{fig:data-collection}, we collect transaction processing times from Etherscan. In this RQ, we statistically analyze the distribution of transaction processing times and determine how gas prices influence these times. More specifically, we classify the gas prices of the studied transactions into five categories, namely: \textit{very cheap}, \textit{cheap}, \textit{regular}, \textit{expensive}, and \textit{very expensive}. The ranges of gas prices for these categories are determined dynamically. More specifically, the gas price category of each studied transaction $t$ is determined by (i) taking the gas price of all transactions included in the 120 blocks that precede the block containing $t$ (see Appendix~\ref{appendix:rq1-blocklookback} for details), (ii) splitting the price distribution into five equal parts (quintiles), (iii) assigning a gas price category to each quintile (first quintile and lower represents \textit{very cheap}, second quintile represents \textit{cheap}, and so on), and finally (iv) mapping the gas price of $t$ to one of the quintiles.

We perform statistical tests to compare multiple distributions. We perform a standard statistical procedure, which consists of running an omnibus test (Kruskal-Wallis), followed by a post-hoc test (Dunn's test) with effect size calculation (Cliff's Delta). By means of this procedure, we can determine whether there is a statistically significant difference between a given pair of distributions (e.g., the processing times associated with \textit{expensive} and \textit{very expensive} price categories) and, in the positive case, determine whether such a difference is meaningful in practice.

More specifically, first we run the Kruskal-Wallis test ($\alpha = 0.05$). The Kruskal-Wallis test is the non-parametric equivalent of the \textit{one-way analysis of variance} (ANOVA). We use the Kruskal-Wallis test to determine whether at least one of the distributions differs from the others with statistical significance. In the positive scenario, we then run the Dunn's post-hoc test ($\alpha = 0.05$) with Bonferroni correction. Dunn's test performs multiple pairwise distribution comparisons. The null hypothesis for each pairwise comparison is that the probability of observing a randomly selected value from the first distribution that is larger than a randomly selected value from the second distribution equals one half (i.e., the same null hypothesis as that of the Wilcoxon signed-rank test). If we observe a statistically significant difference between a given pair of distributions, we calculate the Cliff's Delta ($\delta$) effect size measure in order to better understand the \textit{practical} significance of the difference. Given two distributions $d_1$ and $d_2$, Cliff’s Delta measures how often the values from $d_1$ are larger than those from $d_2$. We assess Cliff's Delta using the following thresholds~\citep{Romano06}: \textit{negligible} for $|\delta| \le 0.147$, \textit{small} for $0.147 < |\delta| \leq 0.33$, \textit{medium} for $0.33 < |\delta| \leq 0.474$, and \textit{large} otherwise.

\begin{figure}
\centering
  \includegraphics[width=0.9\linewidth]{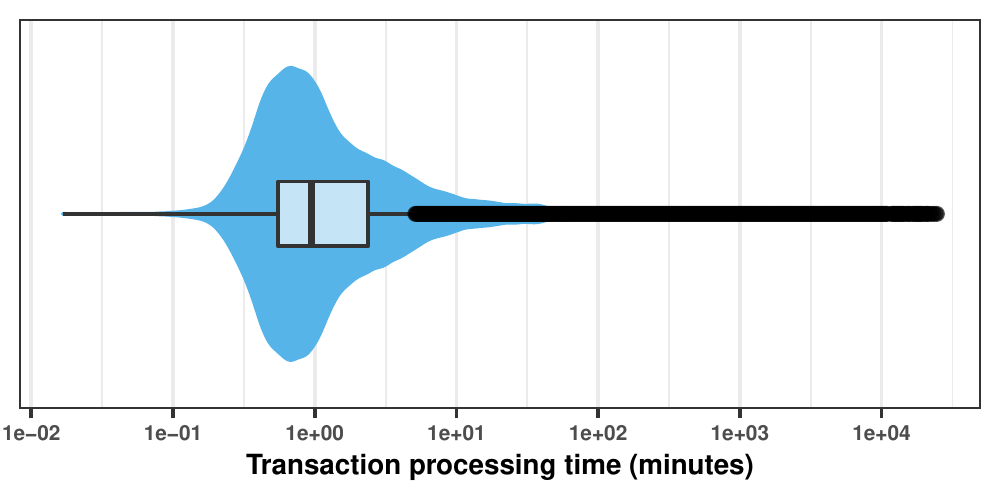}
  \caption{Violin plot depicting transaction processing times (log10 scale).}
  \label{fig:processing_time_dist}
\end{figure}

\smallskip \noindent \textbf{Findings.} \observation{Transactions take a median of 57s to be processed.} Figure \ref{fig:processing_time_dist} depicts the distribution of transaction processing times. Three quarters of the processing times lie in the range of 33s to 2m 23s (interquartile). As another reference point, 90\% of the transactions are processed within 8m. 

\begin{figure}
  \centering
  \includegraphics[width=0.9\linewidth]{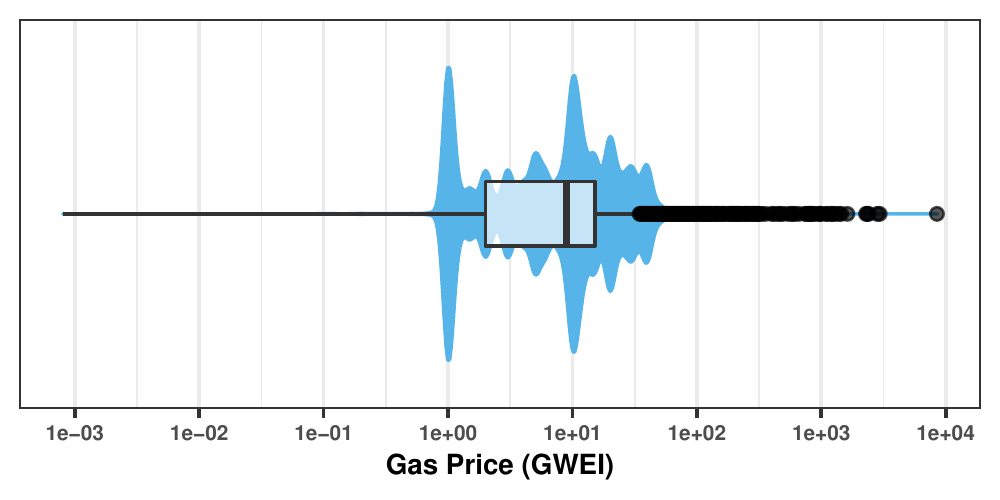}
  \caption{Distribution of gas prices (log10 scale).}
  \label{fig:gas_price_dist}
\end{figure}

\smallskip \observation{Median-priced transactions (9 GWEI) are processed within 3m 04s in 90\% of the cases.} The distribution of gas prices is shown in Figure \ref{fig:gas_price_dist}. Three quarters of the gas prices lie in the range of 2.0 to 15.2 GWEI range (interquartile).

The violin plot suggests that there are specific values for gas price that are recurrently used (e.g., notice the blobs around 1 GWEI and 10 GWEI). Hence, in Table \ref{tab:top5-gas-prices} we show the top-5 most common gas prices and their associated 90th percentile processing time. For instance, 90\% of the transactions priced at 1 GWEI are processed within 56m 41s ($\approx$ 1 hour).

\begin{table}
\centering
\caption{Top-5 most common gas prices set by transaction issuers.}
\label{tab:top5-gas-prices}
\begin{tabular}{rrr}
\hline
  \multicolumn{1}{c}{\textbf{\begin{tabular}[c]{@{}c@{}}Gas Price \\ (GWEI)\end{tabular}}} &
  \multicolumn{1}{c}{\textbf{Occurrence (\%)}} &
  \multicolumn{1}{c}{\textbf{\begin{tabular}[c]{@{}c@{}}90th Percentile \\Processing Time\end{tabular}}} \\ \hline
1  & 13.2 & 56m 39s \\
10 & 9.2  & 03m 08s   \\
20 & 5.3  & 01m 23s  \\
5  & 4.6  & 11m 52s \\
15 & 3.8  & 01m 38s  \\ \hline
\end{tabular}
\end{table}

\smallskip \observation{Very cheap, cheap, regular, expensive, and very expensive are processed within 50m 43s, 05m 36s, 03m 02s, 01m 32s, and 01m 08s respectively in 90\% of the cases.} Table \ref{tab:gas-price-categories} shows gas price statistics for each of the five gas price categories that we defined. Such statistics provide an intuition as to what is considered cheap and expensive in Ethereum during the studied period. Furthermore, Table \ref{tab:gas-price-categories} indicates the 90th percentile processing time for each gas price category. \dapp developers can leverage the information shown in Table \ref{tab:gas-price-categories} to make more informed gas price choices. Figure \ref{fig:appendix-rq1-gas_price_processing_time} in Appendix \ref{appendix:rq1} depicts the distribution of gas prices for each of our gas price categories.

\begin{table}
  \centering
  \caption{Gas price and processing time statistics for each gas price category.}
  \label{tab:gas-price-categories}
  \begin{tabular}{lrrrrrrrr}
  \hline
  \multirow{2}{*}{\textbf{\begin{tabular}[c]{@{}l@{}}Gas Price \\ Category\end{tabular}}} &
    \multicolumn{7}{c}{\textbf{Gas Price (GWEI)}} &
    \multicolumn{1}{c}{\multirow{2}{*}{\textbf{\begin{tabular}[c]{@{}c@{}}90th Percentile\\ Processing Time\end{tabular}}}} \\ \cline{2-8}
   &
    \multicolumn{1}{l}{\textit{Min}} &
    \multicolumn{1}{l}{\textit{Q1}} &
    \multicolumn{1}{l}{\textit{Median}} &
    \multicolumn{1}{l}{\textit{Q3}} &
    \multicolumn{1}{l}{\textit{Max}} &
    \multicolumn{1}{l}{\textit{Mean}} &
    \multicolumn{1}{l}{\textit{Std. Dev.}} &
    \multicolumn{1}{c}{} \\ \hline
  Very Cheap     & 0.0 & 1.0  & 1.0  & 2.4  & 26.6   & 2.6  & 3.5  & 50m 43s \\
  Cheap          & 1.0 & 2.0  & 4.0  & 8.0  & 30.8   & 5.8  & 5.0  & 05m 36s \\
  Regular        & 1.0 & 5.0  & 9.6  & 11.4 & 36.0   & 9.3  & 5.6  & 03m 02s \\
  Expensive      & 1.0 & 10.0 & 12.0 & 19.3 & 40.0   & 14.0 & 6.5  & 01m 32s \\
  Very Expensive & 0.5 & 20.0 & 30.0 & 40.0 & 8559.2 & 32.0 & 55.4 & 01m 08s \\ \hline
  \end{tabular}
\end{table}

\begin{figure}
  \centering
  \includegraphics[width=0.9\linewidth]{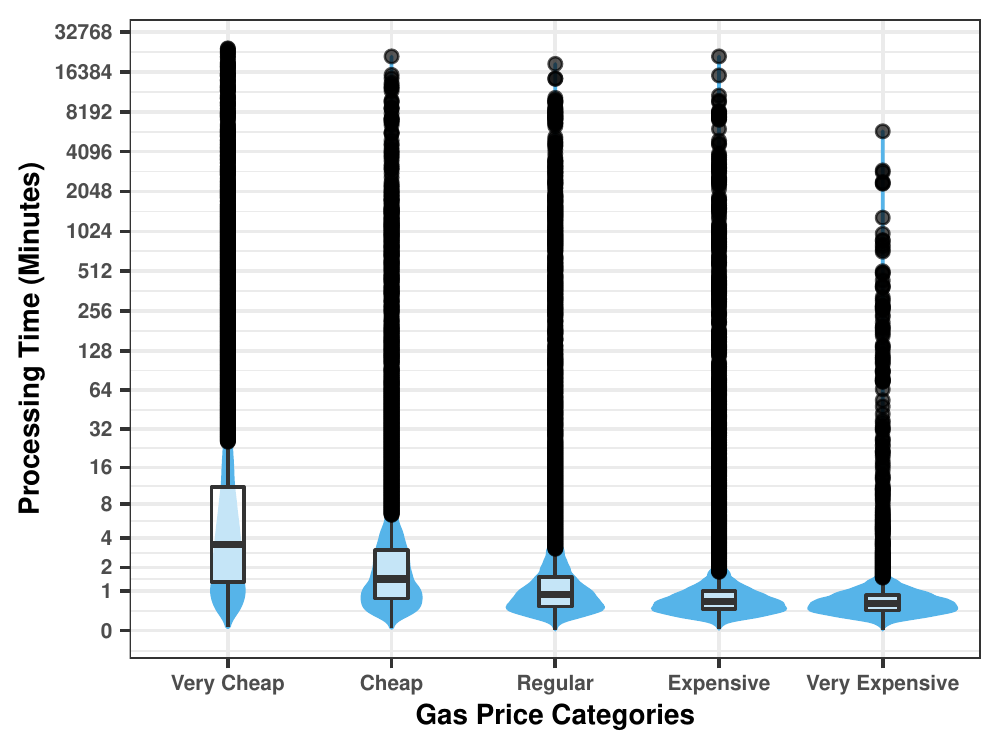}
  \caption{Processing time distribution for each gas price category (log1p scale).}
  \label{fig:gas_price_processing_time}
\end{figure}

\smallskip \observation{Higher gas prices result in faster transaction processing times with diminishing returns.} Figure \ref{fig:gas_price_processing_time} depicts the processing times for each gas price category. The plots indicate that processing times decrease as prices increase (e.g., compare \textit{very expensive} to \textit{very cheap}). Nevertheless, the return over investment diminishes as prices increase (e.g., compare \textit{expensive} to \textit{very expensive}).

In order to further evaluate the relation between gas prices and processing times, we computed a Kruskal-Wallis test ($\alpha = 0.05$) on the distributions shown in Figure \ref{fig:gas_price_processing_time}. The result of the test indicates that at least one of the distributions differs from the others (as we expected from a visual inspection of Figure \ref{fig:gas_price_processing_time}). Next, we perform Dunn's post-hoc test alongside effect size calculations in order to quantify the difference between the distributions. The results for adjacent price categories are summarized in Table \ref{tab:price-vs-txtime}.

\begin{table}
\centering
\caption{Comparison of processing times for pairs of adjacent gas price categories}
\label{tab:price-vs-txtime}
\begin{tabular}{llcr}
\hline
\multirow{2}{*}{\textbf{\begin{tabular}[c]{@{}l@{}}Gas Price\\ Category\end{tabular}}} &
  \multirow{2}{*}{\textbf{\begin{tabular}[c]{@{}l@{}}Other Gas Price \\ Category\end{tabular}}} &
  \multicolumn{2}{c}{\textbf{\begin{tabular}[c]{@{}c@{}}Is there difference in \\ processing times?\end{tabular}}} \\ \cline{3-4} 
           &                & \textit{p.value $\leq$ 0.05} & \multicolumn{1}{c}{\textit{Effect size (delta)}} \\ \hline
Very Cheap & Cheap          & Yes                                & medium (0.38)                                    \\
Cheap      & Regular        & Yes                                & small (0.29)                                     \\
Regular    & Expensive      & Yes                                & small (0.23)                                     \\
Expensive  & Very Expensive & Yes                                & negligible (0.10)                                \\ \hline
\end{tabular}
\end{table}

The results depicted in Table \ref{tab:price-vs-txtime} corroborate our claim of diminishing returns. In particular, despite the statistically significant difference flagged by the Dunn's test, the Cliff's Delta score indicates that the difference in processing times between \textit{expensive} and \textit{very expensive} is actually negligible.

\smallskip \observation{25\% of the very cheap transactions were processed within only 1m 20s.} Since miners' revenue depend on transaction fees, transactions with lower gas prices will receive lower priority and frequently take longer to be processed. Indeed, as depicted in Figure \ref{fig:gas_price_processing_time}, \textit{very cheap} transactions have the highest processing time median. Nevertheless, depending on contextual factors (e.g., the average gas price set by transaction issuers at a given hour of the day), \textit{very cheap} transactions can be processed as fast as \textit{expensive} and \textit{very expensive} transactions. In particular, we observe that 25\% of all \textit{very cheap} transactions were processed within 1m 20s, which happens to be lower than the 90th Percentile processing time for \textit{expensive} transactions. Hence, discovering the moments at which \textit{very cheap} can be processed fast would bring a major financial benefit to \dapp developers. Nevertheless, the inherent variability in the processing time of \textit{very cheap} transactions makes this task far from trivial. In RQ2, we devote particular attention to how accurately online estimation services perform for \textit{very cheap} transactions.

\smallskip \observation{Even very expensive transactions can sometimes take days to be processed.} Although the 90th percentile of the processing time for \textit{very expensive} transactions is only 01m 08s, we note from Figure~\ref{fig:gas_price_processing_time} that some transactions in this price category did take days to be processed. A common reason why very expensive transactions get delayed is the existence of a preceding pending transaction sent by the same transaction issuer, possibly with a cheaper price (please refer to the definition of nonce in Section~\ref{subsec:transactions}). We thus conjecture that the majority of \textit{very expensive} transactions that took days to be processed (i.e., outliers) were waiting for a preceding pending transaction to be processed.

\begin{mybox}{Summary}
  \textbf{RQ1: \RQOne}
  \tcblower
  Transactions are processed in a median of 57s. Also, 90\% of them are processed within 8m. We also observe that:
  \begin{itemize}[topsep = 6pt, itemsep = 3pt, label=\textbullet, wide = 0pt]
    \item Median-priced transactions (9 GWEI) are processed within 3m 04s in 90\% of the cases.
    \item Very cheap, cheap, regular, expensive, and very expensive are processed within 50m 43s, 05m 36s, 03m 02s, 01m 32s, and 01m 08s respectively in 90\% of the cases.
    \item Higher gas prices result in faster transaction processing times with \textit{diminishing} returns.
    \item 25\% of the very cheap transactions are processed within only 1m 20s.
  \end{itemize}
  \textbf{Key takeaways:} \dapp developers can leverage the aforementioned statistics to make more informed gas price choices. \dapp developers would particularly benefit from an online estimation service that produces accurate time estimations for \textit{very cheap} transactions.
\end{mybox}
\subsection{RQ2: \RQTwo}
\label{subsec:rq2}

\noindent \textbf{Motivation.} Given the lack of guarantees regarding transaction processing times in Ethereum, estimators have become a central tool for users of the platform (e.g., \dapp developers). EthGasStation is the most popular transaction processing time estimator. Popular wallets such as Metamask rely on EthGasStation to recommend gas prices for desired processing speeds. Etherscan, one of the most popular Ethereum dashboard, has also recently developed its own processing time estimator. Despite the popularity and relevance of these estimators, their accuracy has never been empirically investigated. 


\medskip \noindent \textbf{Approach.} As outlined in Section \ref{sec:data-collection}, we obtain four processing time predictions, two from EthGasStation (Gas Price API and Prediction Table API) and two from Etherscan (Pending Transaction page and the Gas Tracker). Since we know the actual processing time of each studied transaction, we can determine the accuracy of each of the four models. More specifically, for each studied transaction, we calculate the absolute error ($AE = |{actual - predicted}|$) produced by each model. We then rank the four models based on their AE distributions.

To rank the AE distributions, we follow a three-step approach. In the first step, we statistically analyze the distributions using the same standard approach as that used in RQ1: an omnibus test (Kruskal-Wallis) followed by a post-hoc test (Dunn's test with Bonferroni correction) alongside an effect size calculation (Cliff's Delta). We use $\alpha$ = 0.05 for all tests. In the second step, we define pairwise \textit{win} and \textit{draw} relationships based on the results from Dunn’s test and Cliff’s Delta. We say that one model wins against another when its AE distribution is smaller than that of the other model (Dunn’s p-value < $\alpha$ with a negative non-negligible effect size). Otherwise, we say that there is a draw between the two models (e.g., no statistically significant difference between the two AE distributions). In the third step, we rank the models based on the obtained win and draw relationships. Our ranking adheres to two guiding principles: (i) more wins should make a model rank higher and (ii) winning against a model that has many victories should also make a model rank higher. Fortunately, a direct parallel can be drawn between our guiding principles and the Alpha centrality. Graph centrality measures define the importance of a node in a graph \citep{Newman18}. The Alpha centrality assigns high scores to nodes that have many incoming connections, as well as to those that have an incoming connection from a node that has a high score itself (a recursive algorithm in nature) \citep{Bonacich01}. In other words, the Alpha centrality assigns scores to nodes in a graph based on the idea that incoming connections from high-scoring nodes contribute more to the score of the node in question than equal incoming connections from low-scoring nodes. Alpha centrality is an adaptation of the more popular Eigenvector centrality (the latter has several limitations when applied to directed graphs) \citep{Bonacich01}.

In light of the aforementioned observations regarding Alpha centrality, we rank our models in the third step of our approach as follows: (a) we build a directed graph where nodes represent models, (b) we add an edge with weight 1.0 from model $m_2$ to model $m_1$ when $m_1$ wins against $m_2$ (see the definition of \textit{win}), (c) we add an edge with weight 0.5 from $m_1$ to $m_2$ and vice-versa when there is a \textit{draw} between $m_1$ and $m_2$ (see the definition of \textit{draw}), (d) we calculate the Alpha centrality of each model in the graph\footnote{We use the \texttt{alpha\_centrality} function from the \texttt{igraph} R package (version 1.2.5): \url{https://cran.r-project.org/web/packages/igraph}.}, and (e) we assign ranks to models in accordance to their centrality score (the model with highest centrality score is assigned rank 1, the model with second highest centrality score is assigned rank 2, and so forth). The rationale behind the weights in (b) and (c) is to enforce that wins count more than draws.

Finally, for information purposes, we also summarize the accuracy of the four model in terms of four accuracy measures, namely: Mean Absolute Error (MAE), Median Absolute Error (MedAE), Mean Absolute Percentage Error (MAPE), and Median Absolute Percentage Error (MedAPE).

\smallskip \noindent \textbf{Findings.} \observation{At the global level, the four prediction models are equivalent.} The four models have roughly similar median AEs, which range from 40.8s to 58.2s (see Table \ref{tab:summary-models-rq2} in Appendix \ref{appendix:rq2} for additional performance statistics). We ran our ranking procedure and obtained an identical rank for all models. Upon closer inspection, we observed a statistically significant difference between all pairs of models. Nonetheless, the magnitude of the effect sizes was \textit{negligible} in all cases. Therefore, we consider the four models to have equivalent prediction accuracy at the global level.

\smallskip \observation{When stratifying the analysis by gas prices categories, the Etherscan Gas Tracker webpage outperforms the others for ``very cheap'' and ``cheap'' transactions. In turn, the EthGasStation Gas Price API outperforms the others for the remaining gas price categories.} In RQ1, we observed that there is noticeably more variability in the processing time for \textit{very cheap} and \textit{cheap} transactions (Figure \ref{fig:gas_price_processing_time}). This means that cheaper transactions are inherently harder to predict compared to more expensive transactions. Therefore, we decided to reassess the four prediction models on a per gas-price-category basis (i.e., a stratified analysis). The results that we obtained from running our ranking approach are shown in Table \ref{tab:rq2-models-perf-gas-category}. We use shading to indicate the best performing model(s) for each gas price category.

\begin{table}
  \centering
  \caption{Statistical ranking of the models based on their AEs (per pricing category).}
  \label{tab:rq2-models-perf-gas-category}
  \resizebox{\linewidth}{!}{%
  \begin{tabular}{lcccc}
  \hline
   &
    \textbf{\begin{tabular}[c]{@{}c@{}}Etherscan\\ Gas Tracker Page\end{tabular}} &
    \textbf{\begin{tabular}[c]{@{}c@{}}Etherscan\\ Pending Tx. Page\end{tabular}} &
    \textbf{\begin{tabular}[c]{@{}c@{}}EthGasStation\\ Gas Price API\end{tabular}} &
    \textbf{\begin{tabular}[c]{@{}c@{}}EthGasStation\\ Pred. Table API\end{tabular}} \\ \hline
  \textbf{Very Cheap} &
    \cellcolor[HTML]{EFEFEF}\textbf{1} &
    2 &
    4 &
    3 \\
  \textbf{Cheap} &
    \cellcolor[HTML]{EFEFEF}\textbf{1} &
    2 &
    3 &
    3 \\
  \textbf{Regular} &
    2 &
    4 &
    \cellcolor[HTML]{EFEFEF}\textbf{1} &
    2 \\
  \textbf{Expensive} &
    4 &
    3 &
    \cellcolor[HTML]{EFEFEF}\textbf{1} &
    2 \\
  \textbf{Very Expensive} &
    4 &
    \cellcolor[HTML]{EFEFEF}\textbf{1} &
    \cellcolor[HTML]{EFEFEF}\textbf{1} &
    \cellcolor[HTML]{EFEFEF}\textbf{1} \\ \hline
  \end{tabular}%
  }
\end{table}

Analysis of Table \ref{tab:rq2-models-perf-gas-category} reveals that the Etherscan Gas Tracker webpage outperforms the four other models for the \textit{very cheap} and \textit{cheap} gas price categories. In all remaining price categories, the EthGasStation Gas Price API is the best model (with ties for the \textit{very expensive} category). In other words, if \dapp developers wish to maximize prediction accuracy using the state-of-the-practice models, they would need to use two models that belong to different online estimation services. If developers only care about the prediction of cheaper priced transactions (the hardest ones to predict), then they can rely on the Etherscan Gas Tracker webpage alone. However, we highlight once again that such a model is a black box (see Section \ref{sec:data-collection}). For instance, Etherscan does not provide any public documentation concerning the design of this model and how it operates. A summary of accuracy statistics for the four models on a per gas-price-category basis can be found in Appendix~\ref{appendix:rq2}.


\begin{mybox}{Summary}
  \textbf{RQ2: \RQTwo}
  \tcblower
  From a global perspective, the four studied models are equivalent accuracy-wise (median AEs in the range of 40.8s to 58.2s). In turn, a stratified analysis by gas price category indicates that certain models work best for certain pricing categories. In particular:
   \begin{itemize}[topsep = 6pt, itemsep = 3pt, label=\textbullet, wide = 0pt]
    \item The Etherscan Gas Tracker webpage is the most accurate model for \textit{very cheap} and \textit{cheap} transactions.
    \item The EthGasStation is the most accurate model for \textit{regular}, \textit{expensive}, and \textit{very expensive} transactions.
  \end{itemize}
  \textbf{Key takeaways:} Knowing when cheaper transactions will be processed fast provides a major financial benefit to \dapp developers. Yet, the best performing model for \textit{very cheap} and \textit{cheap} transactions is a black box (Etherscan Gas Tracker).
\end{mybox}

\section{Can a simpler model be derived? A post-hoc study}
\label{sec:post-hoc-study}

\noindent \textbf{Motivation.} In RQ2, we observed that Etherscan Gas Tracker webpage is the most accurate model for predicting the processing times of \textit{very cheap} and \textit{cheap} transactions. However, as of the time this paper was written, Etherscan does not provide any public documentation concerning the design of this model nor how it operates (e.g., the learner being used, the features being used by the model, the model retraining frequency). It is also unclear whether the design of the model has changed since its introduction. We consider that it is undesirable and possibly risky for \dapp developers to have to rely on an estimation service that lacks transparency \citep{Dam18}. Moreover, we observed in RQ2 that the EthGasStation Gas Price API is the most accurate model for the other gas price categories. Hence, maximizing prediction accuracy across gas price categories implies using two models that are provided by different online estimation services. This setting is likely to induce a higher development and maintenance overhead compared to the possibility of relying on a single, inherently interpretable model.

Therefore, in this post-hoc study, we assess our ability to derive an estimator that is simple, inherently interpretable, and yet performs as accurately as a combination of the Etherscan Gas Tracker webpage and the EthGasStation Gas Price API (hereafter called \textit{the state-of-the-practice model}).
 
\smallskip \noindent \textbf{Approach.} In the following, we describe the design and assessment of our model. We organize this description into the following items (a) choice of prediction algorithm, (b) feature engineering, (c) data preprocessing, (d) model validation, (e) accuracy comparison, 

\smallskip \noindent \textbf{a) Choice of prediction algorithm.} We choose a linear regression (ordinary least squares) because it is an \textit{inherently interpretable} model (e.g., as opposed to black box models) \citep{Molnar19}. More specifically, a linear regression model predicts the dependent variable as a weighted sum of the independent variables (features) \citep{Harrell15}. Since there is a clear mathematical explanation for the predicted value of each observation, linear regressions are deemed as inherently interpretable models. We use the \texttt{rms} R package\footnote{\url{https://cran.r-project.org/web/packages/rms} (v5.1-4)}.

\smallskip \noindent \textbf{b) Feature engineering and model specification.} We engineer only one feature which combines recent contextual and historical information of the Ethereum blockchain. Namely, we define this feature as: \textit{the average of the percentage of transactions with gas prices lower than the current transaction in the previous 120 blocks}. 

Our rationale for implementing this feature is that history-based features have yielded positive results in prediction models throughout several software engineering studies \citep{Zimmermann05,Hassan06,Nachi06}. We believe that historical features can also be leveraged in our context. More specifically, we expect that the dynamics of transaction processing are likely to be similar to those of half an hour ago. For instance, if the network was \textit{not} clogged half an hour ago, we expect that it will remain unclogged in the vast majority of cases. In light of this rationale, we designed the \textit{moving average of the percentage of transactions with gas prices lower than the current transaction in the previous 120 blocks} feature. For any given transaction $t$, we take the previous 120 blocks (which cover a timespan of 30 minutes on average) and for each of them calculate the percentage of transactions with gas prices less than $t$, and finally take the average of these percentages. This feature reflects how competitive the gas price for $t$ is, hence if the resulting average percentage is high, the processing time should be fairly quick, and vice versa.

\begin{figure}
    \centering
    \includegraphics[width=0.9\linewidth]{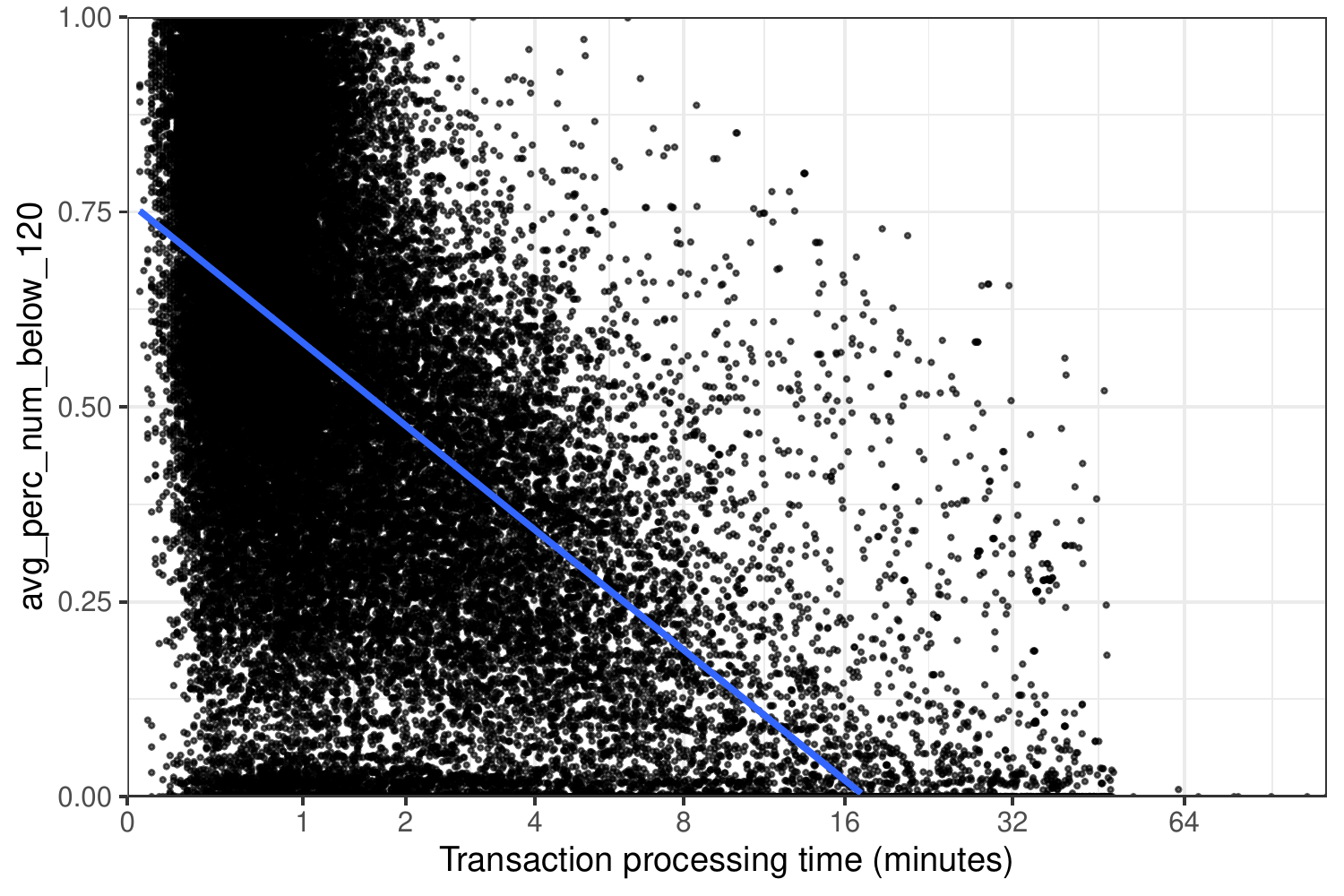}
    \caption{A scatter plot illustrating the relationship between processing time (minutes, log1p scale) and our engineered feature \texttt{avg\_perc\_num\_below\_120}.}
    \label{fig:processing_time-vs-avg_perc_num_below-scatter}
\end{figure}

Next, we investigate the relationship between our engineered feature and the processing times of transactions to ensure our models will have adequate explanatory power of processing times. Figure \ref{fig:processing_time-vs-avg_perc_num_below-scatter} illustrates a negative relationship between the two distributions - as our independent feature increases, processing times decrease, and vice versa. To confirm this we also compute the Spearman's correlation ($\rho$) to analyze the correlation between these two variables. We use Spearman's correlation in lieu of Person's correlation because the latter only assesses linear relationships. Spearman's correlation, in turn, assesses monotonic relationships (linear or not) and is thus more flexible. We assess Spearman's correlation coefficient using the following thresholds \citep{Evans95}, which operate on 2 decimal places only: \textit{very weak} for $|\rho| \leq 0.19$, \textit{weak} for $0.20 \leq |\rho| \leq 0.39$, \textit{moderate} for $0.40 \leq |\rho| \leq 0.59$, \textit{strong} for $0.60 \leq |\rho| \leq 0.79$, and \textit{very strong} for $|\rho| \geq 0.80$. The test results in a score of $\rho$ = -0.55, indicating a moderate, negative relationship between the variables.

\smallskip \noindent \textbf{c) Data preprocessing.} We apply a \textit{log(x+1)} transformation to our feature and dependent variable to cope with skewness in the data. 

\begin{figure}
    \centering
    \includegraphics[width=0.9\linewidth]{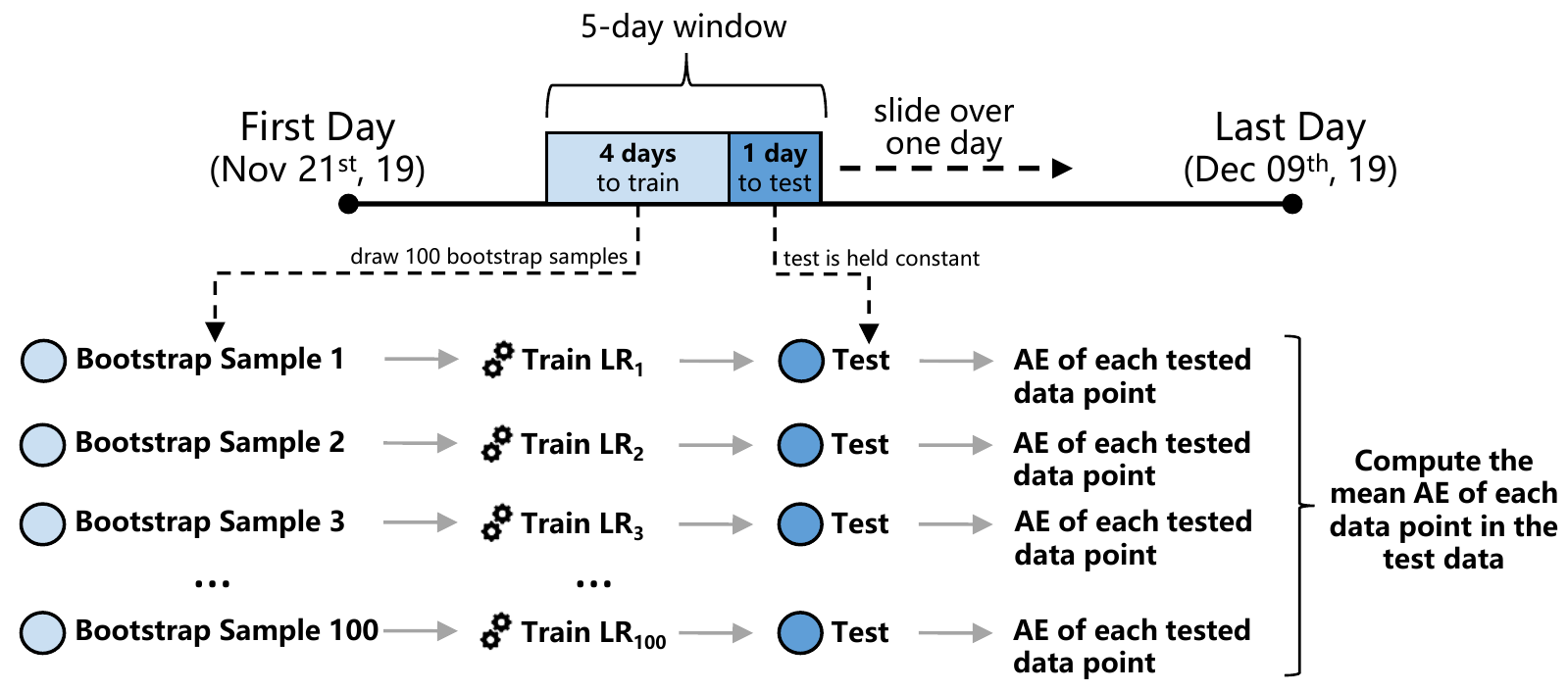}
    \caption{Our sliding-time-window-based model validation approach. LR stands for ``linear regressor'' and AE stands for ``absolute error.''}
    \label{fig:discussion-model-validation}
\end{figure}

\smallskip \noindent \textbf{d) Model validation.} To account for the historical nature of our feature, we employ a sliding-time-window-based model validation approach \citep{Lin18}. A schematic of our model validation approach is depicted in Figure \ref{fig:discussion-model-validation}.

We use a window of five days, where the 
four first days are used for training and the final day is used for testing. We then slide the window one day and repeat the process until the complete period is evaluated. In order to increase the robustness and stability of our model validation, we evaluate each window using 100 bootstrap training samples instead of picking the entire training period only once \citep{tantithamthavorn2017mvt,Efron93}. The test set, in turn, is held constant. Every time we test a model, we compute the absolute error (AE) of each tested data point. After the linear regression model from each bootstrap sample is tested, we compute the mean AE of each data point in the test set. 

\begin{figure}
    \centering
    \includegraphics[width=0.9\linewidth]{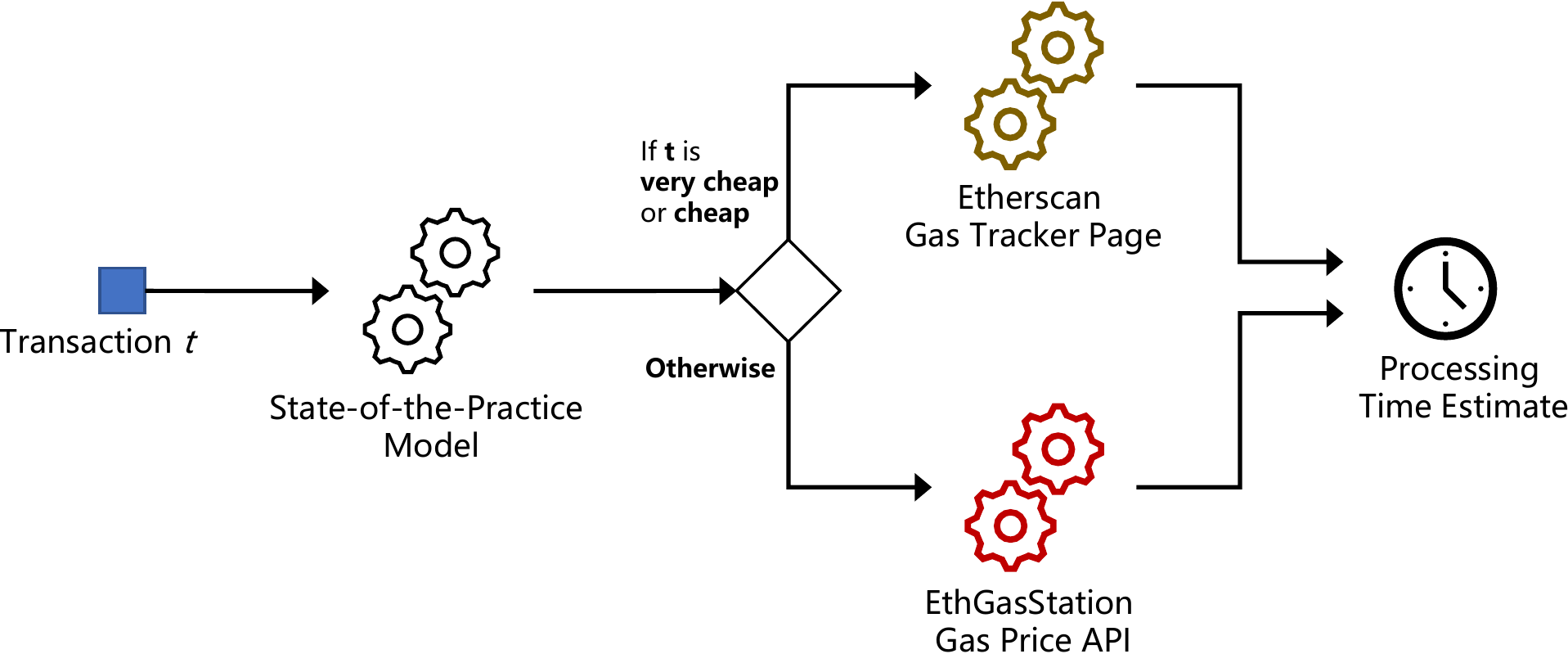}
    \caption{Overview of how the state-of-the-practice model operates.}
    \label{fig:how-competitor-works}
\end{figure}

\smallskip \noindent \textbf{e) Accuracy comparison.} For each transaction in the test set, we first obtain the mean AE given by our model (see Figure~\ref{fig:discussion-model-validation}). Next, we obtain the AE given by the \textit{state-of-the-practice model}. The state-of-the-practice model forwards a transaction $t$ to the Etherscan Gas Tracker webpage when $t$ is either \textit{very cheap} or \textit{cheap}. If $t$ belongs to any other gas price category, the state-of-the-practice model forwards $t$ to the EthGasStation Gas Price API. The state-of-the-practice model can thus be seen as an ensemble that forwards a transaction $t$ of gas price category $c$ to the model that performs best for $c$ (see results from RQ2). Figure \ref{fig:how-competitor-works} illustrates how the state-of-the-practice model operates.

Finally, we investigate whether our model outperforms the state-of-the-practice model. To this end, we compare the AE distribution of our model with that of the state-of-the-practice model. The comparison is operationalized using a Wilcoxon signed-rank test (i.e., a paired, non-parametric test) followed by a calculation and assessment of the Cliff's Delta effect size.

\smallskip \noindent \textbf{f) Potential savings evaluation.} To empirically demonstrate how accurate processing time predictions can lead to money savings (lower transaction fees), we design an experiment to evaluate our models in terms of money saved. To this we 1) draw a sample of blocks from our studied dataset, 2) generate lookup tables using our models, 3) filter artificial transactions using their gas prices and predicted processing times, and 4) verify that actual transactions could have been processed within a faster processing time. We explain each of these steps in detail below.
 

\textbf{Draw a sample of blocks from our studied dataset.} First we draw a statistically representative sample of blocks (95\% confidence level, 5 confidence interval) from all blocks in our studied dataset (see Section \ref{sec:data-collection}).

\textbf{Generate lookup tables using our models.} For each block b in our drawn sample, we then generate artificial transactions for our model to predict on (lookup tables). Each of these transactions possess a gas price starting from 1 GWEI up to the maximum gas price of all processed transactions in that block. 

\textbf{Filter artificial transactions using their gas prices and predicted processing times.} For each transaction $t$ in block $b$ with actual processing time $p$ and gas price $g$, we search for an entry in the lookup table of block $b$ with gas price $g_{target}$ and predicted processing time $p_{target}$, such that $g_{target} < g$ and $p_{target} \leq p$. If multiple entries match the criteria, we sort the matching entries by gas price and pick the one in the middle. We use the gas price $g_{target}$ and predicted processing time $p_{target}$ to determine possible savings for each transaction $t$.

\textbf{Verify that actual transactions could have been processed within a faster processing time.}
Next, we search for factual evidence that setting $t$ with a gas price of $g_{target}$ instead of $g$ could have resulted in $t$ being processed within $p$ (even though $g_{target} < g$). We search for a transaction $t_2$ in $b$ with a gas price $g_2$ such that $g_2 = g_{target}$. If such a transaction does not exist, we cannot verify $g_{target}$ with empirical data. In this case, our experiment outputs “inconclusive”. However, if $t_2$ does exist, we collect its processing time $p_2$. If $p_2 \leq p$, our experiment outputs “saving opportunity” as the gas price of $t$ could have been lower (i.e., a positive result of using our model). We also save both $g_2$ and $g$, such that we can compute how much lower the transaction fee would have been. Conversely, if $p_2 > p$, our experiment outputs “failure to save”, as our models would not have saved users money.

\begin{figure}
    \centering
    \includegraphics[width=0.9\linewidth]{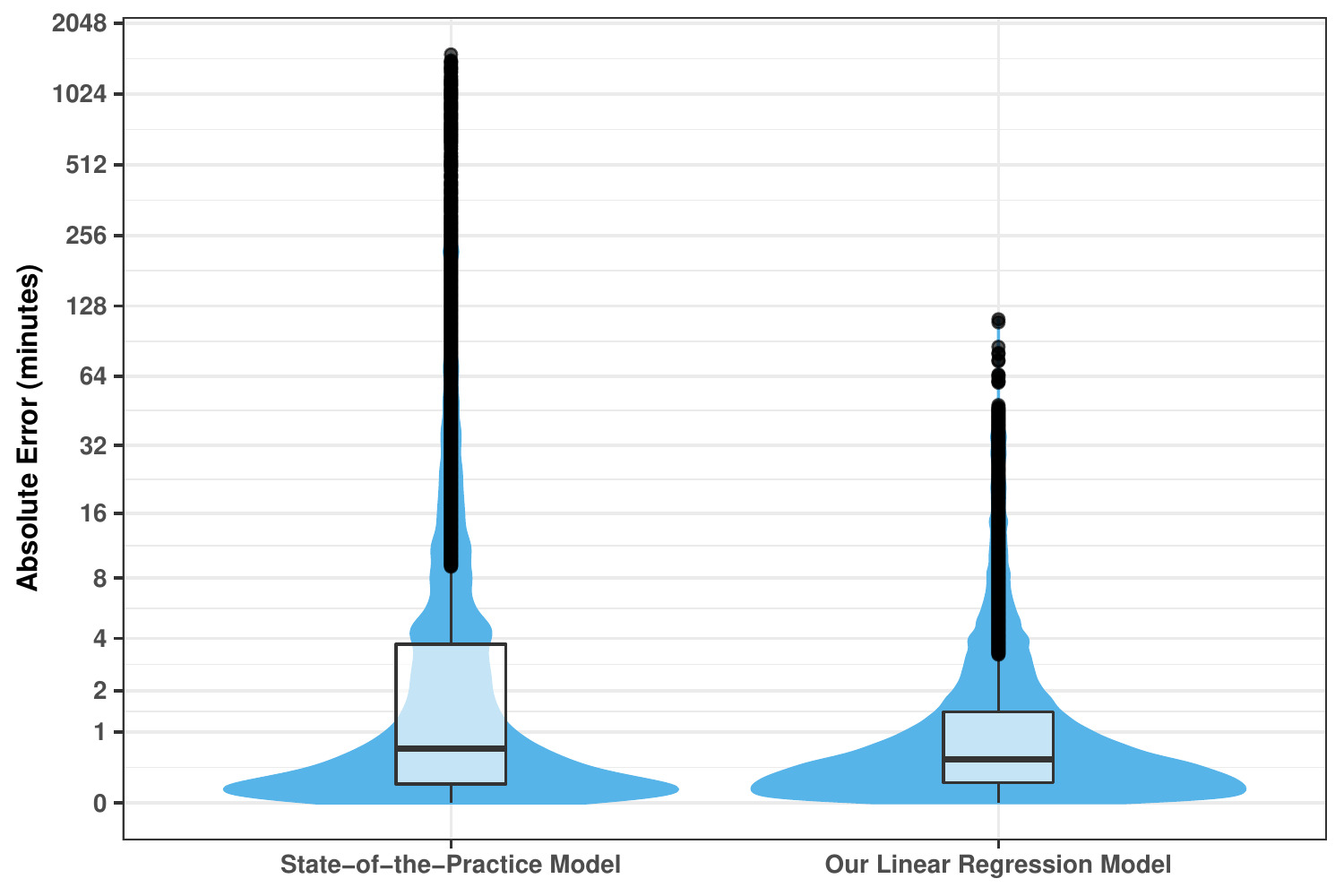}
    \caption{Absolute errors (AEs) of the state-of-the-practice model and ours (lop1p scale).}
    \label{fig:gas-tracker-vs-ours}
\end{figure}

\smallskip \noindent \textbf{Findings.} \observation{At the global level, our model is as accurate as the state-of-the-practice model.} The results are depicted in Figure \ref{fig:gas-tracker-vs-ours}. Analysis of the figure reveals that the distributions have a roughly similar shape. While the medians are also very similar, the third quartile of our model is clearly lower. Indeed, a Wilcoxon signed-rank test ($\alpha = 0.05$) indicates that the difference between the distributions is statistically significant (p-value < 2.2e-16). Nevertheless, Cliff's delta indicates that the difference is negligible ($\delta = |-0.106|$).

\begin{figure}
    \centering
    \includegraphics[width=0.9\linewidth]{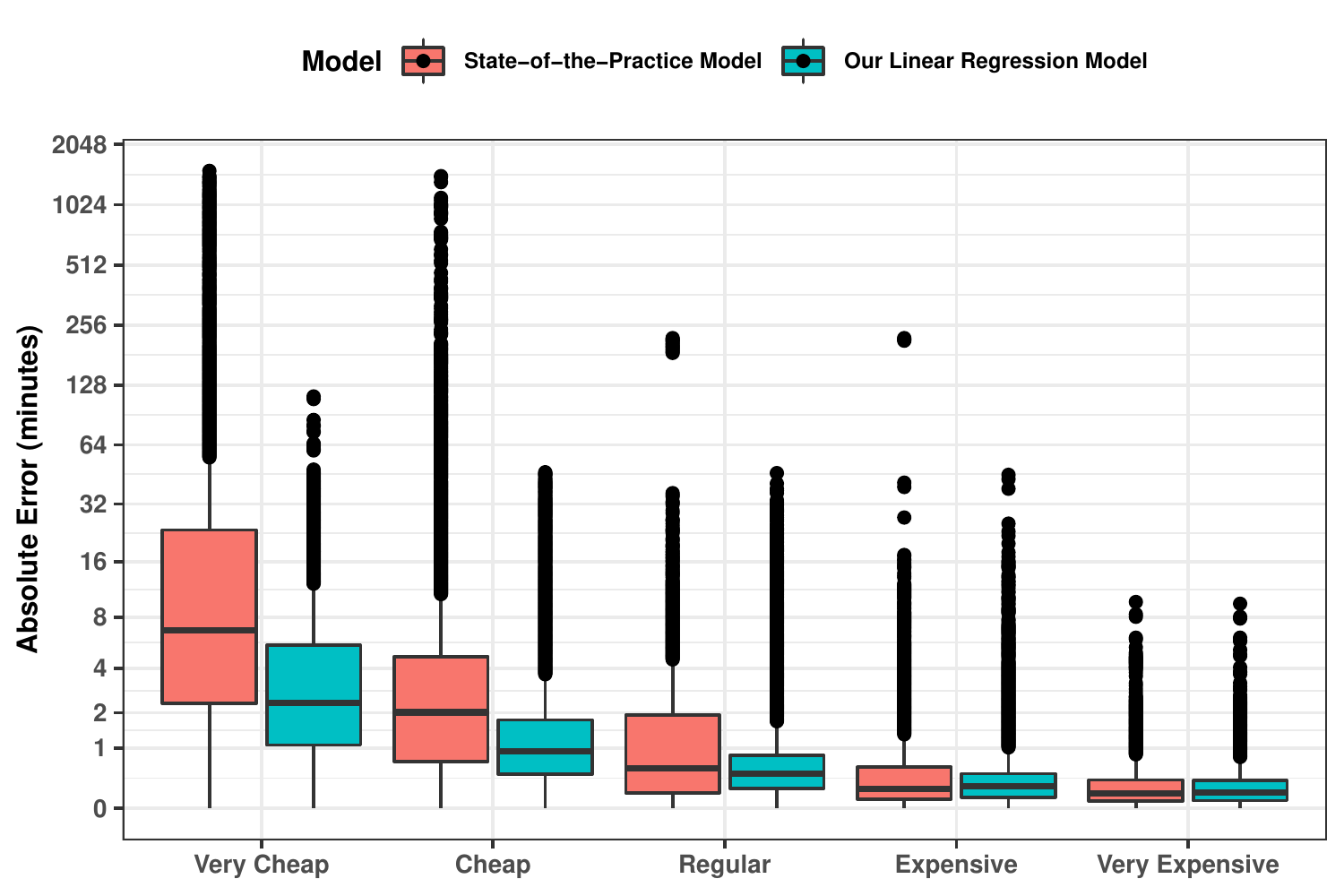}
    \caption{Absolute errors (AEs) of the state-of-the-practice model and ours: stratified by gas price (log1p scale).}
    \label{fig:gas-tracker-vs-ours-per-price-category}
\end{figure}

\smallskip \observation{Our simple linear regression model outperforms the state-of-the-practice model for the ``very cheap'' and ``cheap'' price categories.} Figure \ref{fig:gas-tracker-vs-ours-per-price-category} depicts the results when we split the data according to the gas price categories defined in RQ1. Analysis of the figure seems to indicate that our model outperforms the state-of-the-practice in the \textit{very cheap} gas price category. However in the rest of the categories for (\textit{regular}, \textit{expensive}, and \textit{very expensive} transactions, our model perform just as well as the state-of-the-practice.

To better understand the differences between the two models across gas price categories, we again compute Wilcoxon signed-rank tests ($\alpha = 0.05$) alongside Cliff's Deltas. A summary of the results is shown in Table \ref{tab:summary-results-price-cat}. The results of the statistical tests and effect size calculations show that our simple model performs surprisingly well for cheaper transactions, as it outperforms the state-of-the-practice in both the \textit{very cheap} and \textit{cheap} categories. We recall from RQ1 that transactions in the \textit{very cheap} and \textit{cheap} categories are the most difficult ones to predict, as processing times in these price categories vary considerably more than those in other price categories (Figure \ref{fig:gas_price_processing_time}). Finally, the difference in performance between our model and the state-of-the-practice model is negligible in all remaining price categories. A summary with additional accuracy statistics for our model and the state-of-the-practice can be seen in Appendix \ref{appendix:post-hoc-study}.


\begin{table}
    \centering
    \caption{Summary of the performance difference between our model and the state-of-the-practice model.}
    \label{tab:summary-results-price-cat}
    \resizebox{\textwidth}{!}{%
        \begin{tabular}{lrrcr}
        \hline
        \multicolumn{1}{c}{\multirow{2}{*}{\textbf{Gas Price Category}}} &
        \multicolumn{2}{c}{\textbf{\begin{tabular}[c]{@{}c@{}}Median absolute error\\ (minutes)\end{tabular}}} &
        \multicolumn{2}{c}{\textbf{\begin{tabular}[c]{@{}c@{}}Is there a difference in AE between\\ our model and the state-of-the-practice?\end{tabular}}} \\ \cmidrule(lr){2-3} \cmidrule(lr){4-5}
        \multicolumn{1}{c}{} &
        \multicolumn{1}{c}{\textit{Our Model}} &
        \multicolumn{1}{c}{\textit{State-of-the-practice Model}} &
        \textit{\begin{tabular}[c]{@{}c@{}}p-value \textless 0.05\\ (Wilcoxon test)\end{tabular}} &
        \multicolumn{1}{c}{\textit{\begin{tabular}[c]{@{}c@{}}Cliff's Delta\\ (ours against state-of-the-practice)\end{tabular}}} \\ \cmidrule(lr){1-1} \cmidrule(lr){2-3} \cmidrule(lr){4-5}
        \textbf{Very Cheap}     & 2.35 & 6.73 & YES & medium (-0.38)      \\
        \textbf{Cheap}          & 0.93 & 2.02 & YES & small (-0.32)      \\
        \textbf{Regular}        & 0.49 & 0.59 & YES & negligible (-0.10) \\
        \textbf{Expensive}      & 0.29 & 0.25 & YES & negligible (0.01)       \\
        \textbf{Very Expensive} & 0.20 & 0.19 & YES & negligible (0.01)  \\ \hline
        \end{tabular}%
    }        
\end{table}

\smallskip \observation{Our simple model could have saved users 53.9\% of USD spent in 11.54\% of the transactions in our drawn sample of blocks.} Despite the strong assumption behind our experiment (i.e. that actual processing times correspond to what transaction issuers were originally aiming for), we observe that using our model would have saved users 53.9\% of USD spent in 11.54\% of the transactions from the sampled blocks. Conversely, our model would have failed to save money in 7.31\% of these transactions, and 81.15\% are deemed inconclusive.

\begin{mybox}{Post-hoc study: can a simpler model be derived?}
    Our simple linear regression model performs at least as accurately as a combination of the existing top-performing models (\textit{the state-of-the-practice model}). The only exception is for transactions in the \textit{expensive} gas price category. In particular:
    \begin{itemize}[topsep = 6pt, itemsep = 3pt, label=\textbullet, wide = 0pt]
        \item Our model relies on only one feature, namely: \textit{moving average of the percentage of transactions with gas prices lower than the current transaction in the previous 120 blocks}.
        \item Our model outperforms the state-of-the-practice for \textit{very cheap} and \textit{cheap} transactions.
    \end{itemize}
    \textbf{Key takeaway:} \textit{Very cheap} and \textit{cheap} transactions are the most difficult ones to predict the processing time for (RQ1). Our simple and inherently interpretable model can be used in lieu of the Etherscan Gas Tracker (proprietary, black box) for estimating the processing time of transactions in these price categories.
\end{mybox}
\section{Implications}
\label{sec:implications}

In the following, we discuss the key implications of our findings.

\smallskip \implication{\dapp developers now have a baseline for transactions processing times in Ethereum.} To the best of our knowledge, our study is the first one to empirically investigate transaction processing times in Ethereum. In particular, RQ1 includes several pieces of information that \dapp developers can use as a yardstick. These pieces of information should also be useful for organizations that want to evaluate the suitability of the Ethereum platform as a backend for a prospective \dapp. For instance, in observation 1, we note that three quarters of the processing times lie in the range of 33s to 2m 23s. As part of observation 2, we also show that the most commonly used prices and their respective 90th percentile processing time (Table \ref{tab:top5-gas-prices}). As part of observation 3, we analyze gas prices, group them into categories, and describe the ranges of these categories (Table \ref{tab:gas-price-categories}). These ranges should help \dapp developers understand what is considered cheap and what is considered expensive in terms of gas price. Finally, our statistics for processing time can be used as a baseline to assess the effectiveness of throughput-improving changes in future versions of the Ethereum platform.

\smallskip \implication{\dapp developers can better assess the trade-off between processing time and transaction cost.} In observation 4, we highlight that higher gas prices have diminishing returns in terms of processing time. Therefore \dapp developers should carefully assess whether using higher gas prices is worthy. In particular, our findings indicate that \dapp developers should typically avoid \textit{very expensive} transactions, as there is no practical difference (as per Cliff's Delta) in their processing time compared to \textit{expensive} transactions. 

\smallskip \implication{\dapp developers can make more informed decisions regarding the choice of a prediction model for transaction processing times in Ethereum.} In RQ2, we show that the Etherscan Gas Tracker is the best model for \textit{very cheap} and \textit{cheap} transactions. From RQ1, we know that the processing time of cheaper transactions varies considerably more compared to those of transactions in other price categories. Hence, we believe that \dapp developers will typically benefit the most from using the Etherscan Gas Tracker model. We also show that the EthGasStation Gas Price API is best model for the remaining gas price categories. Therefore, developers who wish to maximize prediction accuracy can combine the two models into an ensemble model (e.g., similarly to our design shown in Figure \ref{fig:how-competitor-works}). Nonetheless, \dapp developers should be aware that, for a given a price category, the best median AEs that can be achieved with these two models are: 7m 7s for \textit{very cheap} transactions, 1m 36s for \textit{cheap transactions}, 28s for \textit{regular transactions}, 14s for \textit{expensive transactions}, and 12s for \textit{very expensive} transactions (Table \ref{tab:summary-models-rq2-per-price}).

In light of the aforementioned results, we conducted a post-hoc study in which we proposed an alternative model that is simple in design and inherently interpretable. We show that our model outperforms the Etherscan Gas Tracker for \textit{very cheap} and \textit{cheap} transactions. In addition, our model performs as accurately as the EthGasStation Gas Price API for the remaining categories. 
Nevertheless, \dapp developers should be the ones making the final call, since the context in which each \dapp operates might be different and might lead to different requirements in terms of prediction accuracy.

\subsection{How about end-users?}

Our paper focuses on the development model where (i) the blockchain complexities (e.g., transaction parameter setup) are hidden from end-users and (ii) developers carry the burden of submitting transactions with proper parameters. In this context, setting up appropriate transaction parameters (e.g., gas price) becomes a clear Software Engineering problem. 

Nonetheless, we acknowledge the existence of several \dapps that follow an alternative development model in which the blockchain complexities are exposed to end-users. In other words, end-users of these DApps have to submit transactions themselves and commonly use a wallet (e.g., Metamask) to do so. When using wallets, end-users can either accept the wallet’s suggested gas prices for different transaction processing speed categories or manually input a gas price. Hence, in this context, end-users also might find it important to achieve a good balance between transaction cost and processing time speed (e.g., in the event where a given end-user frequently uses a certain DApp). In this vein, the two implications discussed in the previous section apply to these end-users as well. Advanced end-users can even use prediction models such as the one that we propose in the post-hoc study in order to make more informed gas price choices.

\begin{mybox}{Lessons Learned}
    \dapp developers should be aware that:
    \begin{itemize}[topsep = 6pt, itemsep = 3pt, label=\textbullet, wide = 0pt]
        \item Higher gas prices come with diminishing returns in terms of transaction processing times.
            \begin{itemize}[wide = 8pt]
                \item DApp developers should likely avoid very expensive transactions, as there is no practical difference (as per Cliff’s Delta) in their processing time compared to expensive transactions.
            \end{itemize}
        \item The processing time of \textit{very cheap} and \textit{cheap} transactions vary significantly more than that of transactions from more expensive categories.
        \begin{itemize}[wide = 8pt,  itemsep = 0pt]
            \item The Etherscan Gas Tracker is the most accurate state-of-the-practice model for estimating the processing time of transactions in those categories.
            \item Our carefully designed yet simple linear regression model outperforms Etherscan’s GasTracker for transactions in those categories (large and small effect sizes respectively). 
            \item Even \textit{very cheap} transactions can be processed fast. For instance, 25\% of the \textit{very cheap} transactions in our dataset were processed within only 1m 20s. Hence, being able to accurately predict the processing time of \textit{very cheap} can lead to major savings in transaction fees.
        \end{itemize}
    \end{itemize}
\end{mybox}

\section{Related Work}
\label{sec:related-work}


\noindent  \textbf{Transaction processing within blockchains.} \citet{Hu} propose a private blockchain to facilitate banking processes, and also focus on the extent of possible impacting factors on transaction processing times. As the authors focus on a private blockchain for the purpose of a single application, a few assumptions about their blockchain preceding their analysis. These assumptions include: ``transaction rates are slower than the real ETH blockchain'', and ``block size is sufficient enough to include all transactions''. Additionally, all experiments are performed using 10 miner nodes and 10 light nodes. By performing experiments within said blockchain, the authors conclude that transaction times are correlated with block generation times. It is also revealed that network congestion does not affect processing times due to the dynamic adjustment of the block difficulty. Also due to this feature, the average generation time of blocks and ultimately transaction processing times are found to be stable regardless of the amount of miner. As the assumptions stated above do not currently apply to the main net Ethereum blockchain, these results cannot be generalized to the main net of Ethereum blockchain.  


\citet{Kasahara} focus primarily on the effect of Bitcoin fee on transaction processing times using queuing theory. In this study they also conclude that transactions with small fees take a much longer time to be processed than those with high fees. This is included by deriving and analyzing the mean processing times of transactions. The authors also separate transactions by their set fee as part of the analysis, though only segregate them into two groups: high and low. The exploration of network congestion is also included, and it is revealed that both high and low Bitcoin fee transactions are impacted heavily as the usage of the Bitcoin network increases. As a possible reason for this is due to the size of blocks in the Bitcoin blockchain being 1 MB, the authors also discuss the implications of increasing the block size. As it is discovered that high fee transactions continue to experience slow confirmation times as high levels of congestion exist after increasing the block size, it is concluded that increasing the block size is not a solution to this issue.

Rather than focusing on variables from transactions and the network, \citet{Rouhani} focus on the differences in transaction processing times within two of the most used Ethereum clients: Geth and Parity. Using a private blockchain, the authors conclude that using the Parity client with the same system configuration as found within the Geth client, transaction processing times experience an 89.8\% increase. Unfortunately, the type of client that was used to execute a transaction is not available within transaction metadata. Additionally, most mining nodes are part of a mining pool, which increases the difficulty of studying the potential impacts of a single mining node. As a result, when conducting analyses on the transactions executed on the blockchain as a whole, this information cannot be inferred.


\medskip \noindent \textbf{Transaction fees.} Instead of analyzing the impact of transaction fees or gas prices on other factors, the work of \citet{Pierro} explores the effect of several different factors on transaction fees within the Ethereum blockchain. The main factors analyzed by the authors include: electricity price, total number of miners, USD to ETH conversion rates, and total pending transactions. The authors discover that both the total number of miners and pending transactions have a large impact on transaction fees.

\citet{Chen} explain how extremely low gas prices can lead to a security vulnerabilities and attacks such as denial of service. It is stated that Ethereum has updated their platform to defend against the possibility of such attacks, however the paper explores if the changes are adequate enough to minimize any disruption within the network. They conclude that the alternative settings are not enough, and that allowing users to set such low gas prices will continue to give malicious users the opportunity to attempt to disrupt the network. The authors propose a solution which dynamically alters the costs of sending transactions as the amount of transaction executions increase. The experimental results of their proposition reveal that it is effective in preventing potential denial of service attacks.

More generally, transaction fees vary depending on the gas price chosen by transaction issuers. Several tools such as Etherscan’s Gas Tracker EthGasStation, and geth provide recommendations for gas prices. Determining whether transaction issuers adopt the exact suggestions issued by those tools is a fruitful research endeavor, as it helps to understand the driving forces behind gas price choices. Gas price choices impact the economics of Ethereum (e.g., the meanings of cheap and expensive gas prices depend on which gas prices are being chosen by transaction issuers at a given point in time), as well as its behavior (e.g., gas prices influence the likelihood that a given transaction will be selected by a miner to be processed).

\smallskip \noindent \textbf{Measurements of the Ethereum network.} \citet{Kim18} argue that, while application-level features of blockchain platforms have been extensively studied, little is known about the underlying peer-to-peer network that is responsible for information propagation and that enables the implementation of consensus protocols. The authors develop an open-source tool called NodeFinder, which can scan and monitor Ethereum’s P2P network. The authors use NodeFinder to perform an exploratory study with the goal of characterizing the network. From data collected throughout the second trimester of 2018, the authors observe that: (a) 48.2\% of the nodes with which NodeFinder was able to negotiate an application session with were considered \textit{non-productive peers}, as these peers either did not run the Ethereum subprotocol or did not operate on the main Ethereum blockchain (or both), (b) 76.6\% of all Mainnet Ethereum peers use Geth and 17\% use Parity, (c) peers using Geth are more likely to adopt the client's latest version compared to Parity, (d) Geth and Parity implementations of a log distance metric yield different results, which calls for an improved standardization and documentation of operations in the protocols (RLPx, DEVp2p, and the Ethereum subprotocol), (e) Ethereum's P2P network size is considerably smaller than that of the Gnutella P2P network, (f) US (43.2\%) and China (12.9\%) host the largest portions of Ethereum Mainnet nodes, and (g) Ethereum nodes operate primarily in cloud environments (as opposed to residential or commercial networks).

In a similar vein, \citet{Silva20} implement and deploy a distributed measurement infrastructure to determine the impact of geo-distribution and mining pools on the efficiency and security of Ethereum. Data was collected from April 1st 2019 to May 2nd 2019 (\textasciitilde 1 month). Some of their key findings include: (a) nodes located in Eastern Asia are more likely to observe new blocks first as several of the prominent pools operate in Asia (nodes in North America are around four times less likely to observe new blocks first), (b) 1.45\% of the mined blocks were empty, (c) a single miner sometimes produces several blocks at the same height, possibly to exploit the uncle block reward system\footnote{Uncle blocks are created in Ethereum when two blocks are mined and submitted to the ledger at about the same time. The one that is not validated is deemed as an uncle block. Miners are rewarded for uncle blocks. Uncle block reward was designed to help less powerful miners.}, and (d) 12 block confirmations might not be enough to deem a given block as final. The authors classify (a) and (b) as \textit{selfish} mining behavior.
 
\citet{Oliva} analyzes transactional activity in Ethereum. They report that, within the time period of their analysis (July 30th 2015 until September 15th 2018), 80\% of the contracts transactions targeted 0.05\% of all smart contracts. This result shows that a minuscule proportion of popular contracts have a large influence over the network. For instance, if transaction issuers happen to submit transactions with higher gas prices for these contracts, then it is likely that the ranges for all price categories will increase (please refer to Section~\ref{subsec:rq1} for an explanation of how we define price categories). This has happened before. For instance, the sudden success of a game called CryptoKitties in late 2017 led to a severe congestion of the entire network and a significant increase in gas prices~\citep{BBC17}.

\smallskip \noindent \textbf{Gas usage estimation.} We differentiate between two types of studies in the field of gas usage estimation: (i) worst-case estimations and (ii) exact estimations.

\smallskip \noindent -- Worst-case estimations: Studies of this type aim to \textit{upper bounds} for the gas usage of a given function (i.e., worst-case gas usage). For example, \citet{Marescotti18} conduct a study in which they focus on discovering the worst-case gas usage of a given contract transaction. The authors introduce the notion of \textit{gas consumption paths} (GCPs), which maps gas consumptions to execution paths of a function. Worst-case gas estimations are derived by analyzing the GCPs of a function with two candidate symbolic methods. The two symbolic methods build on the theory of symbolic bounded model checking~\citep{Biere99} and rely on efficient SMT solvers. The proposed estimation model is only evaluated on a single example contract. Similarly, \citet{Elvira20} introduce a tool called Gasol that takes as input (i) a smart contract , (ii) a selection of a cost model, and (iii) a selected public function, and it infers an upper bound for the gas usage of the selected function. Differently from their prior work \citep{Elvira19}, the cost model of Gasol is vastly configurable. Gasol relies on several tools to extract control flow graphs from smart contracts, which are then decompiled into a high-level representation from which upper bounds can be calculated using a combination of static analyzers and solvers \citep{Elvira08}. Gasol is implemented as an Eclipse plug-in, making it suitable for use during development time. No evaluation of the proposed tool is conducted.

\smallskip \noindent -- Exact estimations: Studies of this type aim to provide exact estimations of gas usage. \citet{Das20} introduce a tool called GasBoX, which takes a smart contract function and an initial gas bound as input (e.g., the gas limit). Next, it determines whether the bound is exact or returns the program location where the virtual machine would run out of gas. The tool is also designed to be efficient, running with complexity in linear time in the size of the program. GasBoX operates by instrumenting the smart contract code with specific instructions that keep track of gas consumption. GasBoX applies a Hoare-logic\footnote{Hoare logic is a formal system with a set of logical rules for reasoning rigorously about the correctness of computer programs.} style gas analysis framework to estimate gas usage. Despite the theoretical soundness of  behind GasBoX, such a tool has three key limitations: (i) it does not account for function arguments and the state of contracts and (ii) it operates on contracts written in a simplified version of Move\footnote{Move is the smart contract programming language under development by Facebook, to be used in their Libra blockchain platform. The language’s definition can be seen at:
\url{https://developers.libra.org/docs/assets/papers/libra-move-a-language-with-programmable-resources/2020-05-26.pdf
}}, which is a programming language in prototypal phase, and (iii) it is evaluated on 13 examples contracts that are written by the authors.  More recently, \citet{Zarir21} proposed a simple historical method for the estimation of gas usage. For a given smart contract function \texttt{f()}, the authors retrieve the gas usage of the prior ten executions of that function and take the mean. The conjecture behind the method is that there are  historical patterns in how functions burn gas and thus the gas usage of a given transaction sent to a function \texttt{f()} should not be too different from that of recent transactions sent to \texttt{f()}. The authors argue that these patterns in gas usage emerge either due to (i) how the contracts are implemented (e.g., their business logic and/or the data that they store) or (ii) how transactions issuers interact with a function (e.g., the amount of data sent as input parameters to the function in question). The proposed estimation method also leverages an empirical discovery: approximately half of the functions that received at least 10 transactions during the Byzantium hard-fork showed an almost constant gas usage. Differently from prior work in the field, the authors conduct a large-scale evaluation of their approach by testing it on all successful contract transactions from the Byzantium hard-fork (\textasciitilde 161.6M transactions). The results indicate that gas usage could be predicted with an RSquared of 0.76 and a median absolute percentage error (APE) of 3.3\%. Based on their findings, they suggest that (i) smart contract developers should provide gas usage information as part of the API documentation and (ii) Etherscan and Metamask should consider providing per-function historical gas usage information (e.g., statistics for recent gas usage).

\smallskip \noindent \textbf{Reducing gas consumption (gas optimization).} \citet{Zou19} conducted semi-structured interviews and an online survey with smart contract developers to uncover the key challenges and opportunities revolving around their activities. According to the authors, the majority of interviewees mentioned that gas usage deserves special attention. In addition, 86.2\% of the survey respondents also declared that they frequently take gas usage into consideration when developing smart contracts. The reason is twofold: ``gas is money'' and ``transaction failure due to insufficient amount of gas''. In terms of challenges, the authors highlight that ``there is a need for source-code-level gas-estimation and optimization tools that consider code readability'', since most gas-optimization tools operate at the bytecode level (e.g., Remix). 

\citet{Chen17} investigated the efficiency of smart contracts by analyzing the bytecode produced by the Solidity compiler. The authors observed that the compiler fails to optimize several gas-costly programming patterns, resulting in higher gas usage (and consequently, higher transaction fees). The authors introduce a tool called GASPER, which can detect several gas costly patterns automatically.

\citet{signer2018gas} studied the gas usage of different parts of a smart contract code by executing transactions with semi-random data. He deployed and executed the transactions in a local simulation of Ethereum blockchain based on the Truffle IDE and \texttt{solc} compiler. With each execution of a transaction, he collected data and mapped to the corresponding section of an abstract syntax tree (AST) of the Solidity source code. He proposed Visualgas, a tool that provides developers with gas usage insights and that directly links to the source code.

More recently, \citet{Tamara20} proposed 25 strategies for code optimization in smart contracts with the goal of reducing gas usage. They developed a prototype, open-source tool called python-solidity-optimizer that implements 10 of these strategies. For 6 of these 10 strategies, the tool can not only detect the code optimization opportunity but also automatically apply it (i.e., change the code). The authors evaluated their tool on 3,018 verified contracts from Etherscan. On average, 1,213 gas units were saved when deploying an optimized contract. Also on average, 123 units were saved for each function invocation. Despite the small savings, the authors argue that hundreds of thousands of dollars are spent on transaction fees daily and thus even small cost savings can sum up to high absolute numbers.
\section{Threats to Validity}
\label{sec:threats}

\medskip \noindent \textbf{Construct Validity}. As we discovered that the block timestamp is not an accurate representation of when the block was exactly appended to the chain (Section \ref{appendix:eval-block-timestamp}), we opt to define the block timestamp as when a new block appears in the \textit{Latest Block} list of Etherscan. This specifically depends on when one of Etherscan's nodes becomes aware of the block being processed. As a result, the processed timestamp that we collect from Etherscan likely has a small, somewhat constant error (lag) compared to the actual, real, processed timestamp. This lag ($\varepsilon$) embeds three elements: (i) the time for Etherscan to become aware of the new block (information propagation in the blockchain P2P network), (ii) the time it takes for Etherscan to update the webpage, and (iii) the time it takes for us to retrieve the information from the webpage. As a result, we train our model with this "lagged" data. The predictions done by our model thus embed such a lag. That is, the predictions of our model are overestimated by a small $\varepsilon$. 

Nevertheless, we emphasize Etherscan is the most popular Ethereum dashboard and numerous tools rely on Etherscan to determine whether a certain transaction $t$ has been mined. If $t$ has been mined, but it hasn't been acknowledged and/or advertised by Etherscan yet, then many would simply consider the transaction not to have been mined. In that sense, we believe that the small $\varepsilon$ in our estimations play a little role in practice. 

Finally, we note that this inaccuracy problem associated with the processed timestamp is not exclusive to our model, as every existing model needs to compute such a timestamp somehow. We conjecture that most of the existing models use the blockchain-recorded timestamp, as it is (i) the most natural choice, (ii) conveniently stored in the blockchain, and (iii) its flaws are not obvious. For instance, since EthGasStation is open source, we managed to inspect its source code and we were able to confirm that it uses the blockchain-recorded timestamp. In fact, we note that EthGasStation even removes negative processing times in their code\footnote{\url{https://github.com/ethgasstation/ethgasstation-backend/blob/master/egs/egs_ref.py\#L256}}.

The Ethereum blockchain does not keep track of the timestamp at which transactions are submitted. Similarly to the block timestamp, we also relied on Etherscan's Pending Transaction Page to discover the \textit{pending timestamp} of a transaction. Therefore, the accuracy of our pending timestamp depends on the accuracy of the data shown by Etherscan.

More generally, this paper employs a best-effort approach to retrieve the pending and processed timestamps of transactions, due to the requirement and challenges involving long term and real time data collection. Future work should investigate the feasibility of devising more robust data collection approaches to collect additional processing time related data over longer periods of time.

In RQ1, we classified gas prices into 5 categories. Our rationale is that gas prices vary substantially (Figure \ref{fig:gas_price_dist}) and transaction issuers (e.g., DApp developers) need to reason about these prices. Reasoning about categories is easier than reasoning about specific prices, especially given that a certain price \textit{x} might be considered high today and then low next week (cryptocurrencies such as Ether are remarkably more volatile than traditional currencies \citep{Long19,coinmarketcap}). For instance, popular applications such as Tripadvisor\footnote{\url{http://www.tripadvisor.com}} and UberEats\footnote{\url{http://www.ubereats.com}} also use an ordinal variable (i.e., \$/\$\$/\$\$\$/\$\$\$\$) to denote how expensive a certain restaurant is. In our study, we segregate prices into the following straightforward categories: \textit{very cheap}, \textit{cheap}, \textit{regular}, \textit{expensive}, and \textit{very expensive}. Nevertheless, any other categorical classification would still be valid provided that it conveys a clear and intuitive separation of prices (e.g., 4 price categories instead of 5).

In RQ2 and in the post-hoc study, we use the \textit{Alpha centrality} measure to rank prediction models. Alpha centrality is an eigenvector-based graph centrality measure. The use of eigenvectors to rank entities based on pair-wise entity relationships (e.g., entity i \textit{wins} over entity j) dates back to the XIX century \citep{Landau1895} (ranking of players in chess tournaments). Therefore, our approach is far from novel and builds on a solid theory. We also note that we used Alpha centrality in lieu of the more popular Scott-Knott ESD technique\footnote{\url{https://github.com/klainfo/ScottKnottESD}} \citep{tantithamthavorn2017mvt} because our data does not fully meet the assumptions of the latter -- and using unsuitable statistical tests and procedures is a common pitfall in software engineering research \citep{Reyes18}. More specifically, both the original Scott-Knott \citep{Scott74} and the ESD variation are inherently parametric and operate on group means in order to form clusters. Our data (absolute error distributions) are long-tailed distributions, with very large outliers. These outliers disturb the clustering process of Scott-Knott and we thus refrain from using it. Nevertheless, we encourage future studies to reinvestigate our results in light of other suitable ranking mechanisms. Indeed, the problem of fair tournament ranking (i.e., devising a fair final ranking of players based on one-on-one match results) is an open research topic, which falls under an umbrella theory called Spectral Ranking \citep{Vigna19}. For instance, recent research advances have been published in the field of operations research (applied mathematics) \citep{Cook2005}.

As part of the Alpha centrality calculation, we build a directed graph using weighted edges. We use an edge weight of 1.0 to denote wins ($w_{win}$) and an edge weight of 0.5 to denote draws ($w_{draw}$). The rationale is that wins should count more than draws. Most importantly, all of our results still hold regardless of the specific choice of $w_{draw}$ provided that $w_{draw} < w_{win}$.

\medskip \noindent \textbf{Internal Validity.} During our data collection process from Etherscan, we employed a data retrieval mechanism that minimizes the amount of requests that are sent towards such dashboard. As a consequence, we do not collect each and every transaction that is sent to the Ethereum blockchain. This may lead to an imbalance in our processing time data. For example, this may result in obtaining processing times that are mostly common, while not retrieving enough processing times of transactions that are processed at exceptionally fast or extremely slow speeds.

Additionally, in RQ1, we conjectured that the majority of very expensive transactions that took days to be processed (i.e., outliers) were waiting for a preceding pending transaction to be processed. Given the aforementioned characteristics of our data collection, we unfortunately cannot verify this conjecture. More specifically, (i) if a given transaction $t_2$ is waiting for some transaction $t_1$ and $t_1$ was submitted before we started our data collection, then our dataset will not include $t_1$ and (ii) if two transactions $t_1$ and $t_2$ are sent during our data collection period, and $t_2$ is waiting for $t_1$, then there is always the chance that our monitor will only capture $t_2$.

\begin{figure}
    \centering
    \includegraphics[width=1.0\linewidth]{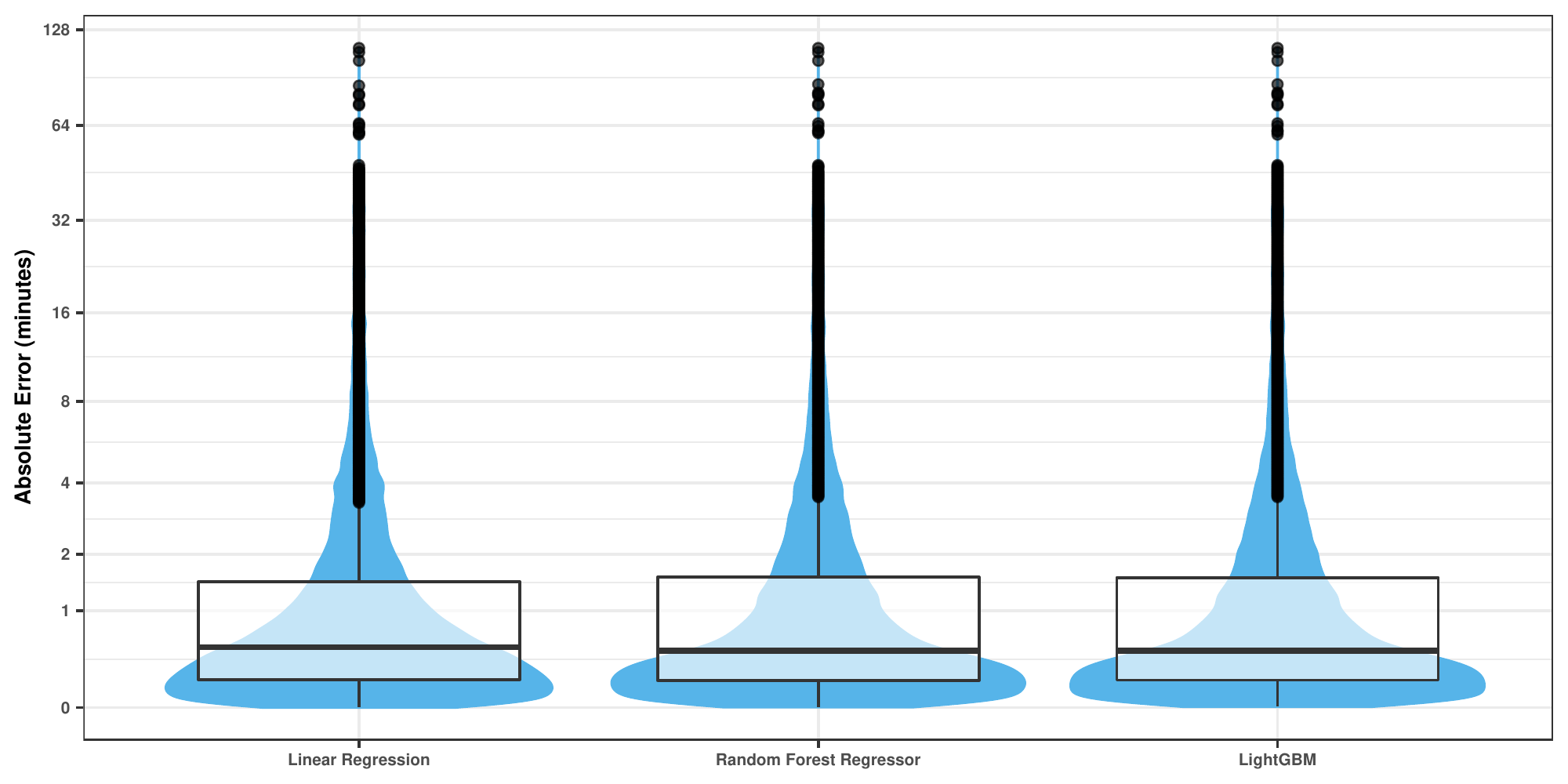}
    \caption{Absolute errors produced by (a) linear regression, (b) Random Forests, and (c) LightGBM.}
    \label{fig:perf-comp-lr-rf-lgbm}
\end{figure}

In the post-hoc study, we built a simple linear regression model. In order to understand how the accuracy of our model compares to that of more sophisticated models, we perform an experiment. We choose two machine learning regressors for this comparison: Random Forests and LightGBM. The former has been extensively used in Software Engineering research with good results~\citep{Ghotra15}. The latter is a complex gradient boosting model that has been used by several top contestants in Kaggle Competitions\footnote{\url{https://github.com/microsoft/LightGBM/blob/master/examples/README.md\#machine-learning-challenge-winning-solutions}}. The results that we obtained are as follows. First, we observe that the models achieve virtually the same performance at the global level (Figure~\ref{fig:perf-comp-lr-rf-lgbm}). Although a Kruskal-Wallis test ($\alpha$ = 0.05) indicates that at least one the performance distributions differs from the others, computation of Cliff’s Delta reveals a negligible difference between any given pair of distributions.

\begin{figure}
    \centering
    \includegraphics[width=1.0\linewidth]{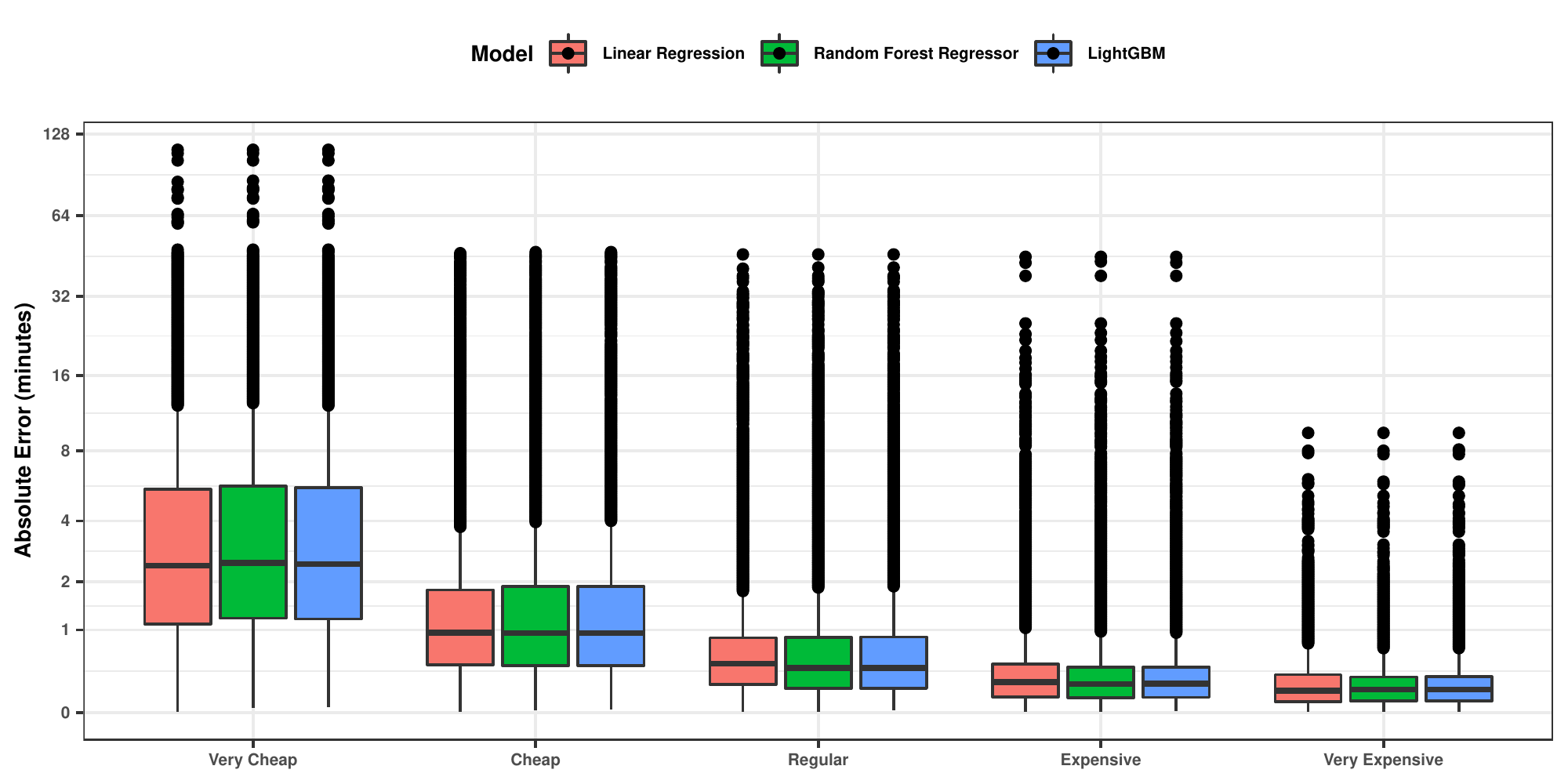}
    \caption{Absolute errors per price category produced by (a) linear regression, (b) Random Forests, and (c) LightGBM.}
    \label{fig:perf-comp-gascat-lr-rf-lightgbm}
\end{figure}

Next, we evaluated how the models perform for different price categories. We employ the same ranking approach described in the approach of RQ2. The distributions are shown in Figure~\ref{fig:perf-comp-gascat-lr-rf-lightgbm}. We observe that all models rank the same in all price categories. Consequently, we conclude that a linear regression is the best choice, since it is simple and inherently interpretable~\citep{Molnar19}.

Finally, in our study we observe that 0.05\% of transactions in our dataset are processed very quickly (e.g., 5 seconds or lower). In these cases, the relative error in our measurements of the prediction accuracy of our models might be high. 


\medskip \noindent \textbf{External Validity}. Our study analyzed data from Ethereum during a specific time window (November 21st 2019 to December 09th 2019). Our results do not generalize to the entire Ethereum transactional history. The main reason is that Ethereum has undergone drastic changes in workload over time (e.g., heavy network congestions during late 2017 due to the boom of the CryptoKitties game \citep{BBC17}). For instance, if our study captured data during the CryptoKitties boom, our results would be bound to a time where the gas prices differ from the majority of gas prices in periods before and after the boom. Hence, regardless of the specific time window chosen to study, it is always possible that the contextual factors of the Ethereum blockchain will change in the future, making generalization extremely difficult. Other contextual factors such as the market capitalization of Ether, the ETH to USD exchange rate, and market speculation also influence how much transaction issuers pay for gas, which then influences transaction processing times. Finally, changes in the protocol introduced in the several hard-forks of Ethereum can also play a role in transaction processing times. We emphasize that a key goal of our paper is to introduce a general and extensible approach for collecting and analyzing transaction processing times (as the actual results per se can always vary across different time frames).

\begin{figure}
    \centering
    \includegraphics[width=1.0\linewidth]{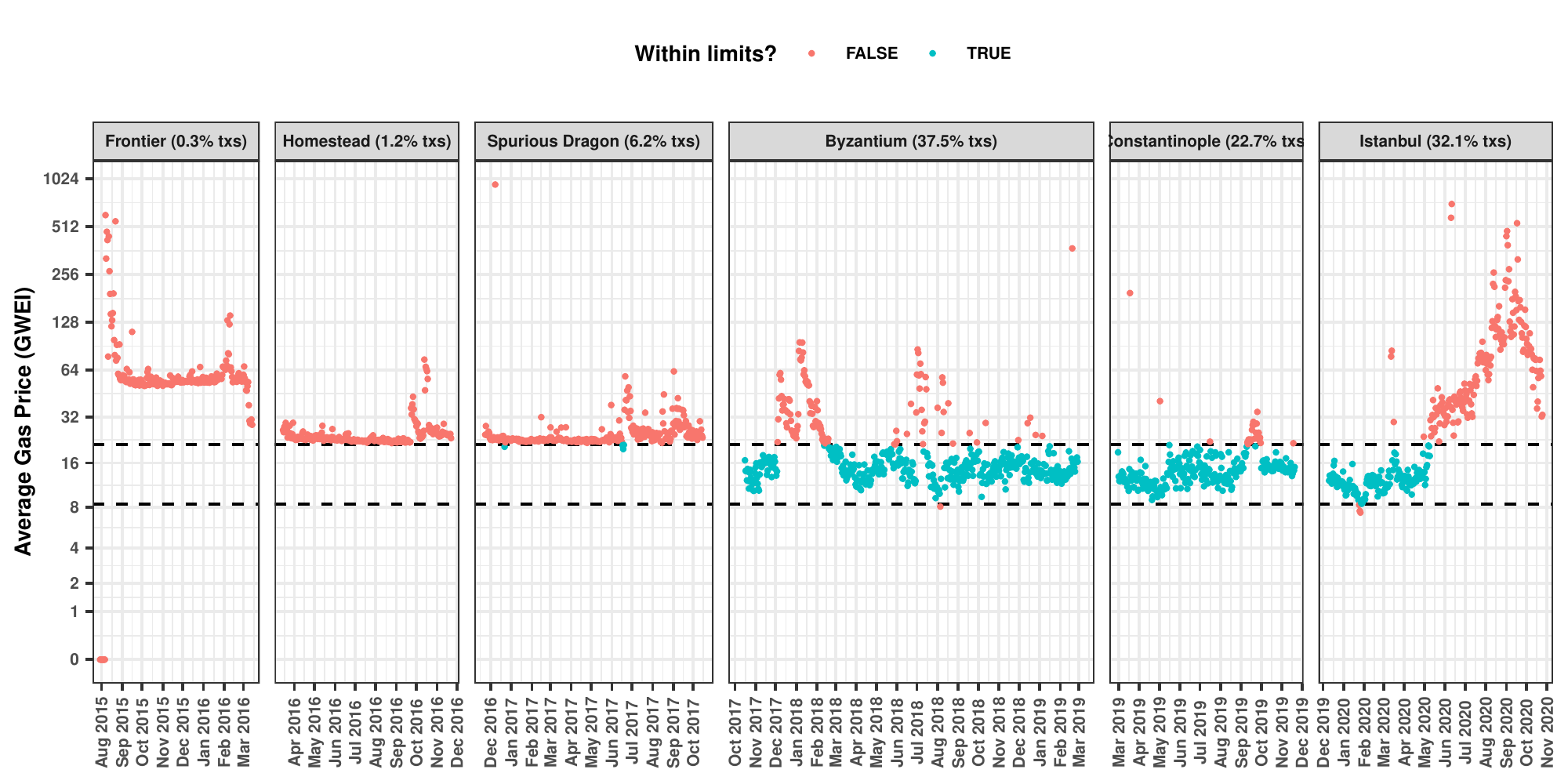}
    \caption{Representativeness of the transactions within our analyzed window. The black dashed lines delineate an interval, which corresponds to \textit{mean gas price} $\pm$ \textit{2 std. deviations} computed over the transactions within our analyzed window. Green dots fall within the interval, whereas red dots do not.}
    \label{fig:avg-gas-price-over-time}
\end{figure}

Despite all the aforementioned factors, it still makes sense to determine how \textit{representative} our window of transactions is. To tackle this problem, we conducted an experiment in which we analyze how the gas prices of transactions inside our analyzed window compare to the gas prices of transactions outside our analyzed window (both past and future). Our rationale is that gas prices serve as a key indicator of the workload state of the network (e.g., EthGasStation). By observing how representative the gas prices of our analyzed window are, we can infer how representative our investigated transactions are (i.e., determine whether they differ from the norm). More specifically, we (i) retrieved all the transactions from our analyzed window, (ii) computed the mean gas price over all such transactions, (iii) defined a gas price boundary corresponding to \textit{mean gas price} $\pm$ \textit{2 std. deviations}, (iv) calculated the average gas price per day from the first day of Ethereum (July 30th 2015) until October 25th 2020, and (v) determined how often these daily averages fall within the aforementioned gas price boundary. The results that we obtained are depicted in Figure~\ref{fig:avg-gas-price-over-time}. 

Analysis of the figure reveals that transactional activity is concentrated in the three last hard forks (note the percentage beside each hard fork name). When focusing on these three hard forks, we note that 71\% of the daily averages fall within our gas price boundary (blue dots). Most of the red dots occur in late Istanbul (May 2020 onwards). The remarkable increase in gas prices is associated with (i) network congestion caused by the boom of the DeFi market (decentralized finance) and (ii) speculation around the release of Ethereum 2.0 \citep{Cointelegraph20a,Cointelegraph20b}. However, a careful analysis of the red dots within the Istanbul period reveals that prices went up, reached a plateau, and are now decreasing. 
In summary, apart from the unusual, recent surge in gas prices, we conclude that our window of analyzed transactions is a reasonable representation of the normal transactional history of Ethereum in terms of gas prices.

Since the selected time frame will always play a role, we encourage future studies to reuse our study design to analyze different, and possibly larger, time frames.
Also, although we choose to analyze gas prices to verify the representativeness of our data as they are known to have a large impact on processing times, there are other features (including those unused by our models), which might have similar levels of predictive power. More generally, future work should extensively evaluate the predictive power of several factors that have the potential to impact transaction processing time, such as gas prices, transaction prioritization algorithms employed by miners, ETH market capitalization (e.g., as proxified by the ETH to USD exchange rate), network hash power (i.e., processing power of the network), and network workload (e.g., number of transactions in the pending pool). Furthermore, our study analyzed data from a specific blockchain platform (Ethereum) during a specific time frame. Hence, our results are unlikely to generalize to other blockchain platforms.

The majority of our study heavily relies on the data provided by Etherscan, which is considered as one of the earliest Ethereum projects aiming to provide equitable access to blockchain data\footnote{\url{https://etherscan.io/aboutus}}. Although we perform an evaluation of the accuracy of the collected pending timestamps, it is possible that these timestamps are still prone to inaccuracies due to the nature of the P2P characteristics of the network. In particular, the geolocation, propagation delay, and network congestion could affect these timestamps by delaying exactly when nodes from Etherscan first see pending transactions. In turn, we encourage future work to perform a detailed analysis on the impact of geolocation on Etherscan’s measured timestamps (e.g., sending transactions from nodes across different geolocations over long periods of time).


\medskip \noindent \textbf{Conclusion validity}. The conclusions that we draw in this paper derive directly from the employment of parametric statistical tests and the computation of Cliff’s Delta effect size measures. This is standard practice in Empirical Software Engineering research. In practice, however, DApp developers might use different criteria to compare the accuracy of two candidate models. For instance, the median absolute errors of our proposed model are higher than those of the state-of-the-practice model for transactions in the expensive and very expensive gas price categories (Table 7). Practitioners may thus choose the state-of-the-practice model for these price categories if they believe that the median absolute error is a more suitable indicator for their use-case compared to Cliff’s Delta.
\section{Conclusion}
\label{sec:conclusion}

Transactions are at the forefront on how information is exchanged on the blockchain. Yet, it is generally unclear how long transactions commonly take to be processed in Ethereum. Predicting the processing time of transactions is key to development of cost-effective \dapps, since developers need to optimize the balance between cost (transaction fees) and user-experience (transaction processing speed). Few online services exist to help smart contract developers estimate how long transactions will take to be processed. Most importantly, the accuracy of these estimation services remain unclear.

In this paper, we collected data from Etherscan, EthGasStation, and Google BigQuery in order to empirically determine how long transactions tend to take in Ethereum. We also evaluate the estimation accuracy of the models employed by the state-of-the-practice estimation services (Etherscan and EthGasStation). Our results led us to conclude that (i) \dapp developers should typically avoid \textit{very expensive} transactions, as there is no practical difference in their processing time compared to \textit{expensive} transactions, (ii) the state-of-the-practice services are far from perfect, leaving considerable room for improvement (especially for cheaper transactions), and (iii) our simple linear regression model outperforms those services for \textit{very cheap} and \textit{cheap transactions}.

We hope that our findings will encourage researchers to conduct further empirical studies in this important area. In particular, we see a clear need to address two interrelated challenges: (i) determine the extent to which contextual factors influence transaction processing times (e.g., network congestion) and (ii) design more accurate processing time estimation services. We also hope that researchers can leverage our study design and supplementary package to bootstrap investigations of the two aforementioned challenges. Finally, we believe that investigating transaction processing times under different settings (e.g., other time frames) and using different techniques (e.g., alternative model ranking approaches) is also a fruitful follow up to our study.
\section*{Disclaimer}
Any opinions, findings, and conclusions, or recommendations expressed in this material are those of the author(s) and do not reflect the views of Huawei.
\bibliographystyle{ACM-Reference-Format}
\bibliography{references}

\appendix
\newpage
\section{Computing Transaction Processing Times}
\label{appendix:tx-processing-time}

Computing the processing time of a given transaction depends on obtaining the \textit{pending timestamp} and the \textit{processed timestamp} (Figure~\ref{fig:txn_lifecycle}). However, obtaining accurate values for these timestamps is considerably challenging. 

In the following, we describe the approaches that we employed to obtain and evaluate the accuracy of the \textit{pending timestamp} (Section~\ref{appendix:pending-timestamp}) and the \textit{processed timestamp} (Section~\ref{appendix:processed-timestamp}).

\subsection{Pending timestamp}
\label{appendix:pending-timestamp}

Despite the ledger nature of a blockchain, Ethereum does not record any data regarding the timestamp at which a transaction was first seen in the network (i.e., \textit{pending timestamp}). Discovering such a timestamp is challenging. First, as described in Section~\ref{subsec:transactions}, each miner node has its own pending pool (i.e., there is no unified, centralized pending pool). Second, the pending pool of a given miner is rarely exposed to the outside world. Third, even if we were to set up our own nodes in the network, there are only so many nodes that we would be able to deploy. Given the peer-to-peer architecture of the blockchain network and the associated broadcasting of transactions, our nodes would likely take a long time to become aware of pending transactions and thus our obtained \textit{pending timestamps} would not be accurate. As a reference, Ethermine\footnote{\url{https://ethermine.org}}, one of the largest Ethereum mining pools, has more than 300k nodes distributed across the globe as of October 2020.

In Section~\ref{appendix:obtaining-pending-timestamp}, we describe how we overcome the aforementioned challenges to obtain the \textit{pending timestamp} of transactions. In Section~\ref{appendix:evaluating-pending-timestamp}, we evaluate the accuracy of our obtained timestamps.

\subsubsection{Obtaining the pending timestamp}
\label{appendix:obtaining-pending-timestamp}

\begin{figure}
    \centering
    \includegraphics[width=1.0\linewidth]{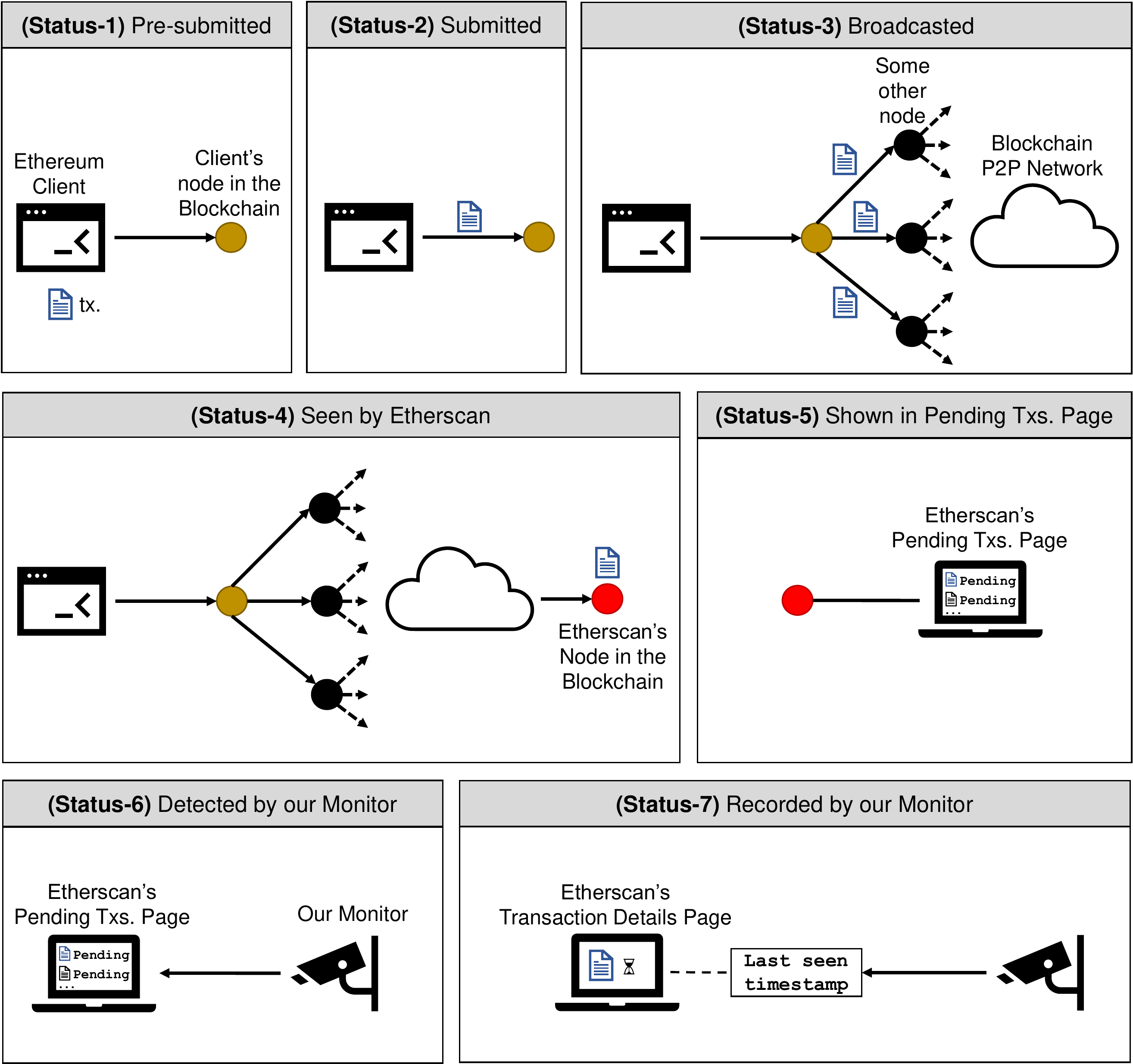}
    \caption{Our approach to detect the pending timestamp of a transaction.}
    \label{fig:appendix-pending-timestamp}
\end{figure}

We obtain an approximation of the \textit{pending timestamp} of a transaction. More specifically, we rely on Etherscan and equate the \textit{pending timestamp} of a transaction \textit{t} with the instant at which an Etherscan’s node first sees \textit{t} (instead of the instant at which \textit{t} is first seen in the network). Our approach is summarized in Figure~\ref{fig:appendix-pending-timestamp}. The figure highlights the various statuses that a transaction \textit{t} undergoes until we can retrieve its \textit{pending timestamp}. In the following, we describe each step.

\begin{figure}[H]
    \centering
    \includegraphics[width=1.0\linewidth]{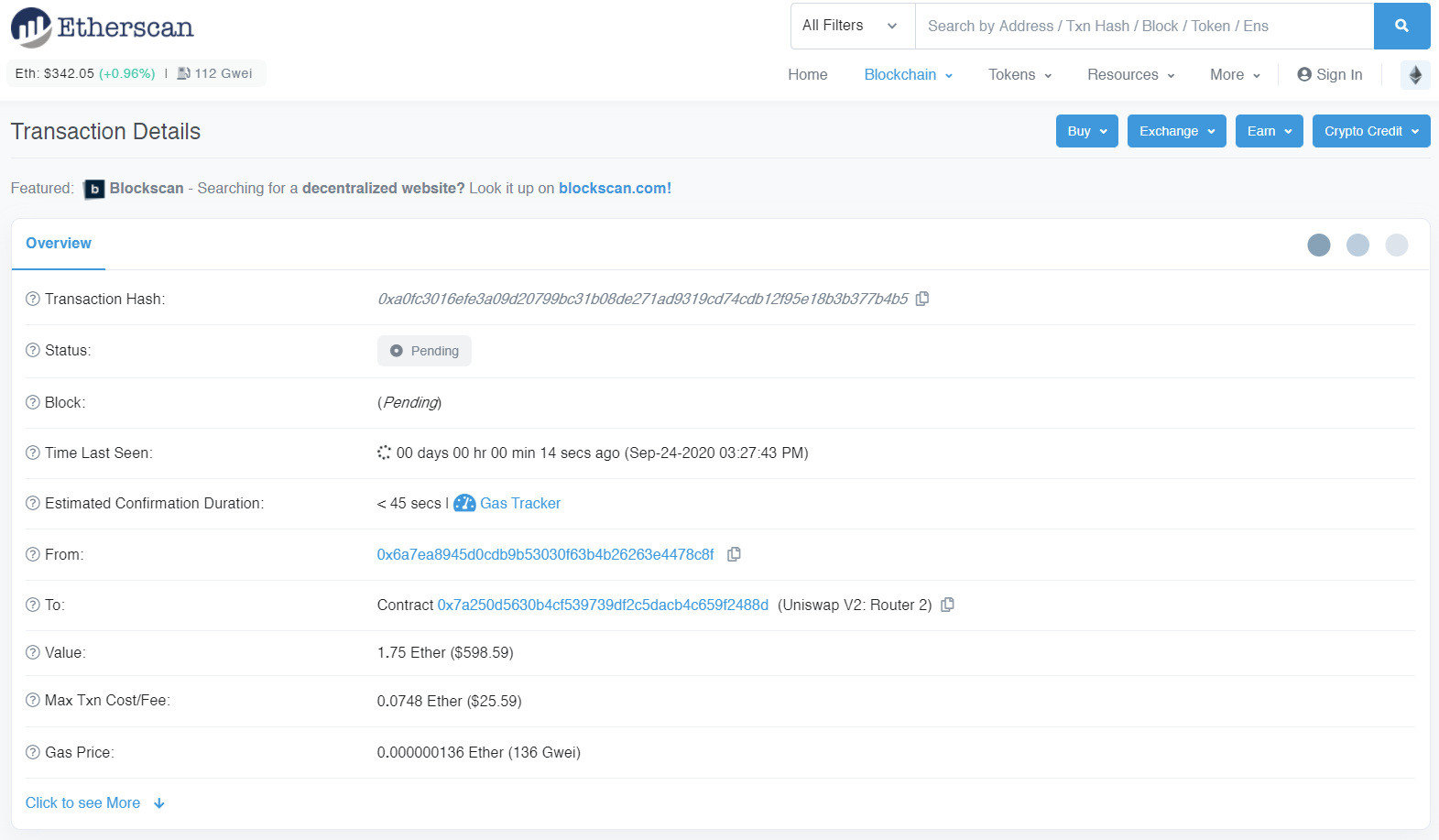}
    \caption{The transaction details webpage of a real-world pending transaction in Ethereum.}
    \label{fig:appendix-tx-details}
\end{figure}

\begin{description}[topsep = 6pt, itemsep = 3pt, wide = 0pt]
    \item[(Status-1) Pre-submitted.] First, the transaction issuer builds the transaction \textit{t} using an Ethereum client tool. We note that the Ethereum client is connected to the transaction issuer’s node in the Ethereum peer-to-peer network. 
    
    \item[(Status-2) Submitted.] The transaction is submitted by the Ethereum client. 
    
    \item[(Status 3) Broadcasted.] The issuer’s node broadcasts the transaction to its neighbour nodes, which in turn broadcast the transaction to their neighbour nodes and so on. 
    
    \item[(Status 4) Seen by Etherscan.] The transaction eventually reaches some node belonging to Etherscan. This is the instant at which Etherscan becomes aware of the transaction \textit{t}. From the perspective of Etherscan, such an instant corresponds to the \textit{pending timestamp}. 
    
    \item[(Status 5) Shown in Pending Txs. Page.] Some Etherscan node (likely the one that first saw the pending transaction \textit{t}) communicates with the Etherscan’s \textit{Pending Transactions} webpage\footnote{The url of the Pending Transactions webpage is \url{https://etherscan.io/txsPending}. There is no public information regarding how Etherscan's nodes in the network and Etherscan's Pending Transactions page communicate with each other (Etherscan is not open source). We conjecture that a Publish-Subscriber model is implemented.} to signal the existence of \textit{t}. The pending transaction \textit{t} is then added to the list of pending transactions in the \textit{Pending Transactions} webpage.
    
    \item[(Status 6) Detected by our Monitor.] We built a monitor that watches the Pending Transactions webpage for updates. Our monitor detects \textit{t} in the table of pending transactions.
    
    \item[(Status 7) Recorded by our Monitor.] Our monitor accesses the \textit{Transaction Details} webpage associated with the pending transaction \textit{t}. This webpage includes a field called ``Time Last Seen'', which contains two pieces of information: (a) a live ``stopwatch'' that increases second by second and (b) the \textit{pending timestamp} in parenthesis. We record the \textit{pending timestamp}. In Figure~\ref{fig:appendix-tx-details}, we display the transaction details page of a real-world transaction \textit{t}.
\end{description}

\begin{figure}[H]
    \centering
    \includegraphics[width=1.0\linewidth]{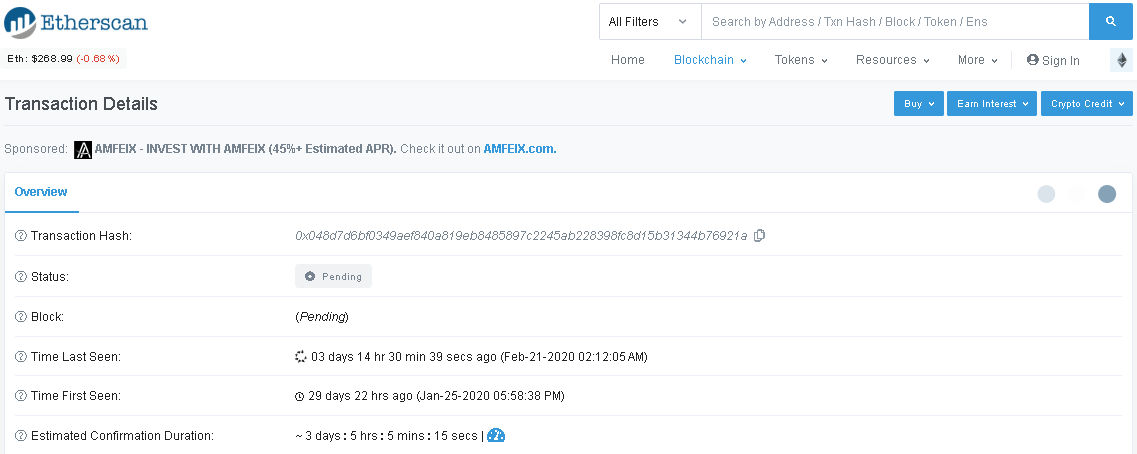}
    \caption{The transaction details webpage of a real-world transaction that has been pending for more than 29 days. Note the addition of the ``Time First Seen'' field in comparison to Figure~\ref{fig:appendix-tx-details}.}
    \label{fig:appendix-tx-firstseen-lastseen}
\end{figure}

When transactions take a long time to be processed (days in our practical experience), Etherscan changes the transaction details page. More specifically, a new field called ``Time First Seen'' is added and the ``Time Last Seen'' field is updated (Figure~\ref{fig:appendix-tx-firstseen-lastseen}). Although Etherscan does not publish any documentation describing these fields in detail, we manually observed that (i) the ``Time First Seen'' is a fixed timestamp, (ii) such a timestamp is older than that of the ``Time Last Seen'' (the portion in parenthesis), and (iii) the ``Time Last Seen'' stopwatch resets to zero and restarts. The new meaning of the ``Time Last Seen'' field is not clear to us. Since the ``Time First Seen'' field contains an older timestamp, we use that as the \textit{pending timestamp} when such a field is shown.

\subsubsection{Evaluating the accuracy of the collected pending timestamps}
\label{appendix:evaluating-pending-timestamp}

We designed an experiment to evaluate the accuracy of our collected \textit{pending timestamps}. In a nutshell, we send transactions to Ethereum, record the timestamp at which they were sent (\textit{submitted timestamp}), record their \textit{pending timestamp}, and compare these two timestamps for every transaction. Our goal is to determine the delta between the two timestamps and to determine whether (i) the delta is large and (ii) whether the delta changes much from transaction to transaction.

The detailed experiment design is as follows. We set up a program to submit transactions. Since transaction processing times vary based on gas price, we decided to submit transactions using various prices. More specifically, our program sends five transactions every hour, one in each of the following gas price categories: \textit{very cheap}, \textit{cheap}, \textit{regular}, \textit{expensive}, and \textit{very expensive}. As explained in Section~\ref{subsec:rq1}, the ranges of gas prices for these categories are determined dynamically using a quintile approach over the gas prices of transactions residing in the 120 most recent mined blocks. We began the experiment on November 20, 2019 and executed it for 40 hours straight. Hence, a total of 200 transactions were sent (40 in each gas price category). 
 
Right before sending a transaction \textit{t}, the program saves a \textit{sent timestamp}. Next, the \textit{pending timestamp} for that transaction is obtained using the approach described in Figure~\ref{fig:appendix-pending-timestamp}. Once the timestamp is obtained and recorded, the program proceeds to send the next transaction in the next gas price category. Finally, we compare the \textit{sent timestamp} with the \textit{pending timestamp} of each transaction.

\begin{figure}[H]
    \centering
    \includegraphics[width=1.0\linewidth]{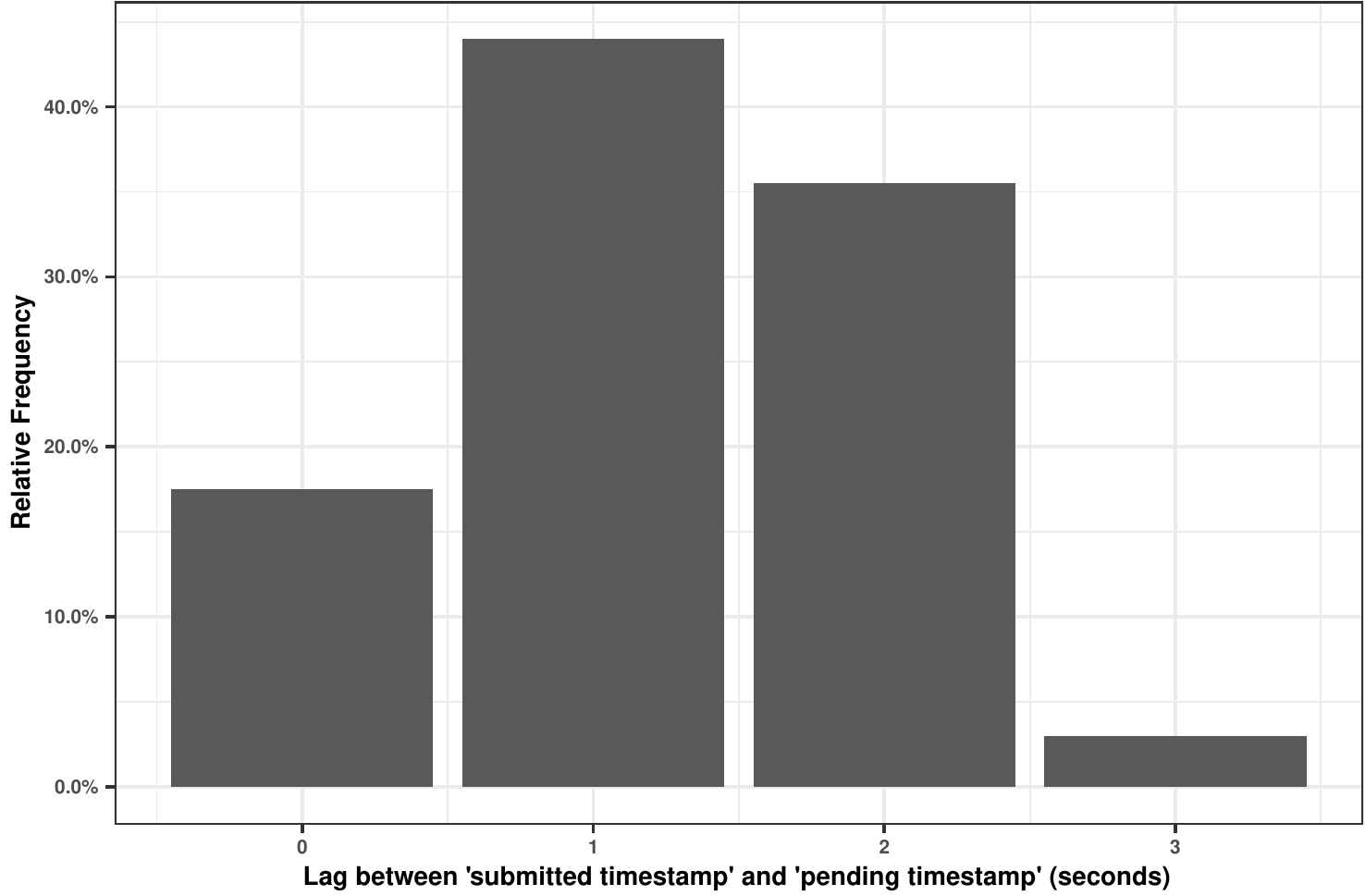}
    \caption{Lag (delta) between \textit{submitted timestamp} and \textit{pending timestamp} for our 200 submitted transactions.}
    \label{fig:appendix-lag-between-sub-and-pend}
\end{figure}

\smallskip \noindent \textbf{Results.} \textbf{Etherscan becomes aware of pending transactions in 1 to 2 seconds in 79.5\% of the cases.} The results that we obtained are depicted in Figure~\ref{fig:appendix-lag-between-sub-and-pend}. Analysis of the figure reveals that the lag between the \textit{submitted timestamp} and the \textit{pending timestamp} is small (3 seconds at most) and stable (in the range of 1 to 2 seconds in 79.5\% of the cases). Indeed, since Etherscan is a real-time dashboard of Ethereum, Etherscan needs to have ``many'' nodes in the network in order to quickly and accurately capture the state of the network. Consequently, we conclude that the \textit{pending timestamps} that we collected using the approach described in Section~\ref{appendix:obtaining-pending-timestamp} are a good approximation of the original, unknown \textit{pending timestamps} of transactions.

\subsection{Processed timestamp}
\label{appendix:processed-timestamp}

The \textit{block timestamp} is recorded in the blockchain and indicates the timestamp at which a given block has been appended to the blockchain. The \textit{processed timestamp} of all transactions inside a mined block \textit{b} correspond to the \textit{block timestamp} of \textit{b}. However, the \textit{block timestamp} is potentially inaccurate by construction. Due to the peer-to-peer architecture of the blockchain network, there is no global clock. Miners not only have to rely on their own clock, but might also have their own algorithm to decide upon the exact time for the block timestamp. This timestamp is accepted as long as the timestamp of the mined block is greater than the timestamp of the parent block and no longer than 2 hours into the future \citep{wood2019ethereum}. Therefore, block timestamps can drift considerably from miner to miner.

In Section~\ref{appendix:eval-block-timestamp}, we empirically evaluate the accuracy of the blockchain-recorded \textit{block timestamps}. We conclude that these timestamps are inaccurate. In Section~\ref{appendix:obtain-better-block-timestamp}, we describe our approach for retrieving more accurate \textit{block timestamps}.

\subsubsection{Evaluating the accuracy of the blockchain-recorded block timestamps}
\label{appendix:eval-block-timestamp}

We investigated the \textit{block timestamps} of the mined blocks that ended up housing the transactions that we submitted as part of the experiment described in Section~\ref{appendix:evaluating-pending-timestamp}. More specifically, we calculated the delta between the \textit{block timestamp} and \textit{submitted timestamp} for our 200 submitted transactions and analyzed the distribution of deltas. 

\begin{figure}[H]
    \centering
    \includegraphics[width=1.0\linewidth]{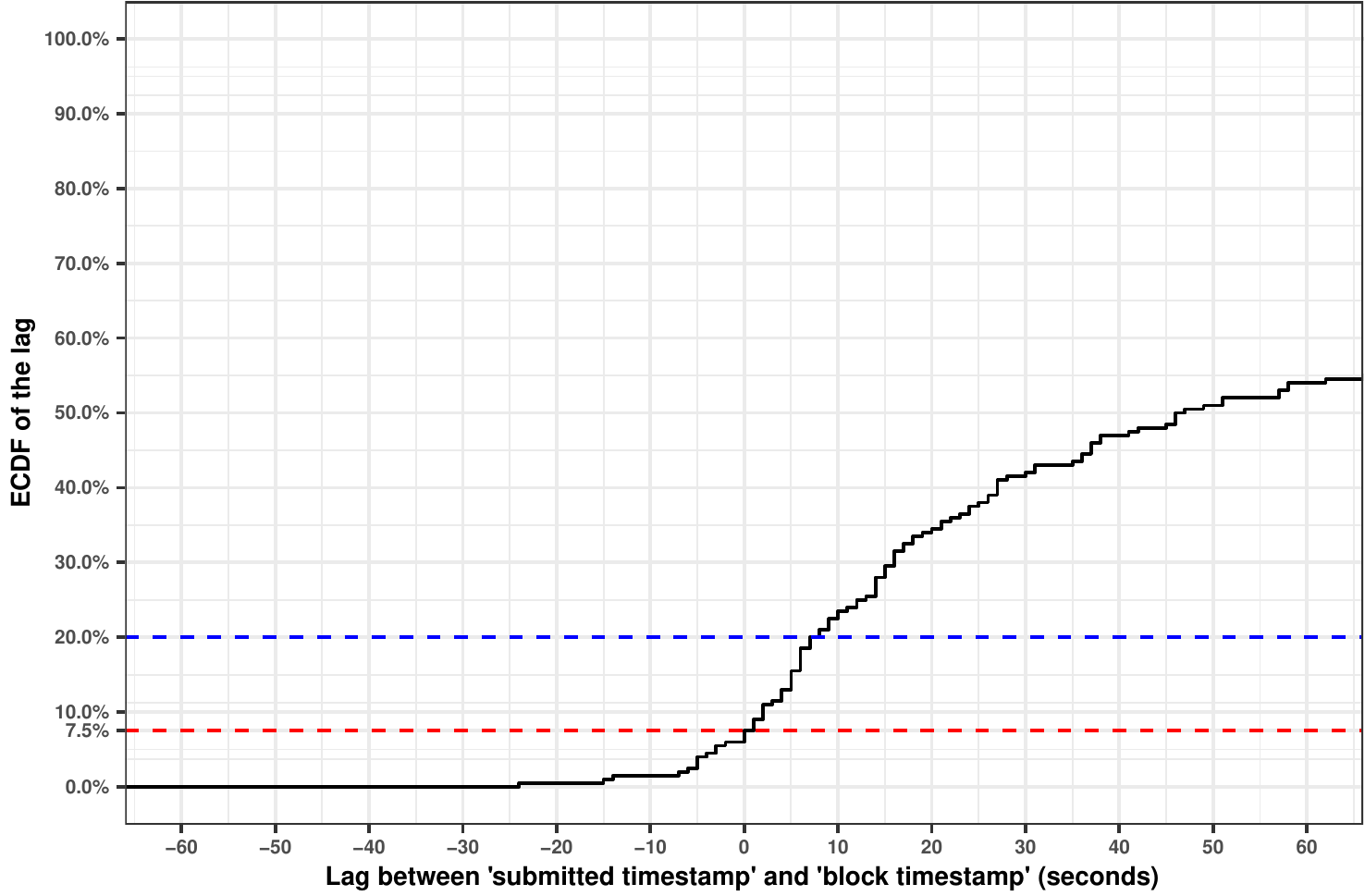}
    \caption{Lag (delta) between \textit{submitted timestamp} and \textit{block timestamp} for our 200 submitted transactions.}
    \label{fig:appendix-lag-between-sub-and-block}
\end{figure}

\smallskip \noindent \textbf{Result. The block timestamp is inaccurate, as such a timestamp happened to be lower than or equal to the transaction submission timestamp in 7.5\% of the cases.} The results that we obtained are shown in Figure~\ref{fig:appendix-lag-between-sub-and-block}. As the red dashed line indicates, 7.5\% of our submitted transaction had either a negative or a zero lag between their submission time and the blockchain-recorded block timestamp. For instance, in one of our submitted transactions, the \textit{submitted timestamp} recorded by our program was \texttt{2019-11-11 20:55:58 UTC}. In turn, the block timestamp associated with the transaction was \texttt{2019-11-11 20:55:56 UTC}\footnote{\url{https://etherscan.io/tx/0xb9d4eb05900e1f455b13bc671d4e6d36576cbe047251714860096b8141ce2611}}.

This provides empirical evidence that the block timestamp, as recorded in the blockchain, is indeed inaccurate. As another reference, the red blue dashed line indicates that 20\% of our submitted transactions were processed in no more than 7 seconds. Given that blocks are appended 15s in average, we believe that it is unlikely that 20\% of our transactions would be processed in no more than 7 seconds. Therefore, we conclude that the block timestamp is an inappropriate proxy for the \textit{processed timestamp}.

\subsubsection{Obtaining a more accurate block timestamp}
\label{appendix:obtain-better-block-timestamp}

\begin{figure}[H]
    \centering
    \includegraphics[width=1.0\linewidth]{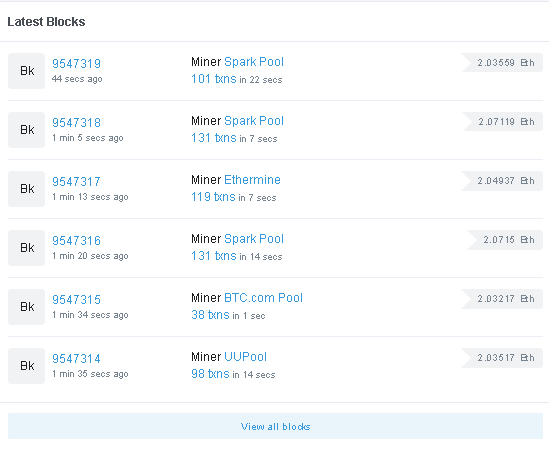}
    \caption{Live list of the latest blocks that were appended to Ethereum (Etherscan’s front page).}
    \label{fig:appendix-latest-blocks}
\end{figure}

To obtain a more accurate value for the processed timestamp, we use another piece of information provided by Etherscan. More specifically, Etherscan triggers an update on its front page when it discovers that a new block has been appended to the blockchain. Each new block is shown at the top of a live, real-time list of newly appended blocks (Figure~\ref{fig:appendix-latest-blocks}).

Therefore, instead of relying on the clocks of several different miners, we monitor Etherscan’s front page and record the timestamp at which a new block appears at the top of the live list. We use this timestamp as the processed timestamp. By employing such an approach, our only source of bias is now Etherscan’s method for updating this live list. Since Etherscan aims to provide a real-time dashboard, we believe that the lag between the addition of a new block and Etherscan’s corresponding page update is minimal and somewhat constant across blocks.

\begin{mybox}{Summary}
    \begin{itemize}[itemsep = 3pt, label=\textbullet, wide = 0pt]
        \item Transaction processing time is calculated as the delta between the \textit{pending timestamp} and the \textit{processed timestamp}.
        \item We obtain the \textit{pending timestamp} from Etherscan's pending pool page.
        \item The \textit{block timestamp} is miner-specific and our experiment shows that it is too inaccurate to be used as proxy for the \textit{processed timestamp}.
        \item We obtain the \textit{processed timestamp} from Etherscan's front page (live list of new blocks).
    \end{itemize}
\end{mybox}
\newpage
\section{RQ1: Gas price distribution for each gas price category}
\label{appendix:rq1}

\begin{figure}[H]
  \centering
  \includegraphics[width=\linewidth]{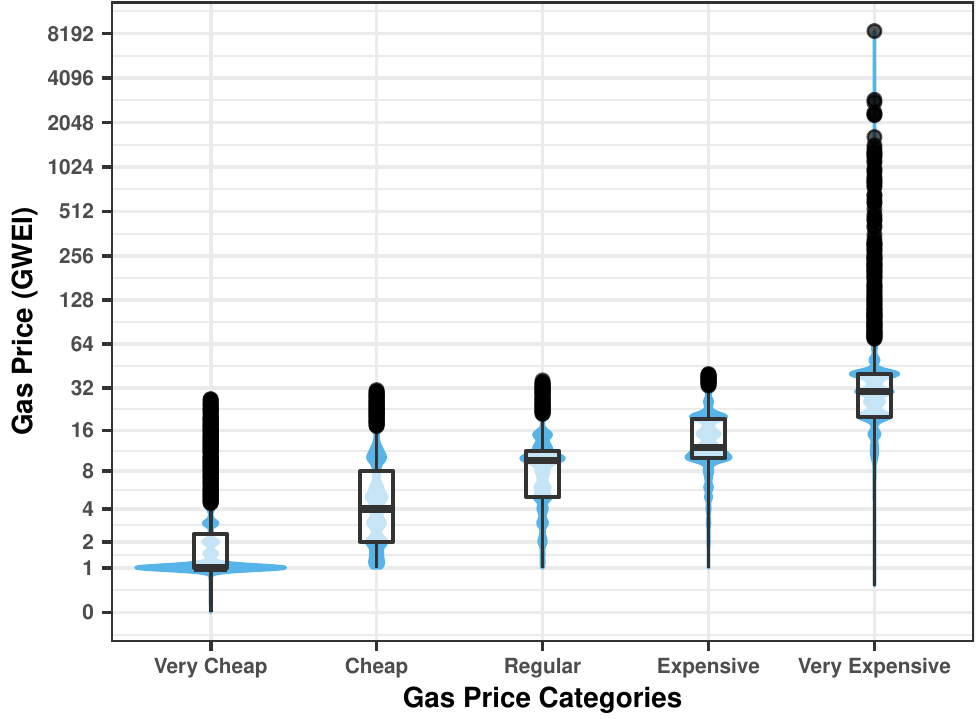}
  \caption{Gas price distribution for each gas price category.}
  \label{fig:appendix-rq1-gas_price_processing_time}
\end{figure}

The gas price distribution for each gas price category is shown in Figure \ref{fig:appendix-rq1-gas_price_processing_time}.

\subsection{Sensitivity Analysis on Block Lookback}
\label{appendix:rq1-blocklookback}

In Ethereum, a new block is appended every 15 seconds on average and each block can contain a limited number of transactions. Therefore, the gas price of transactions is strongly influenced by a demand-supply relationship. In other words, to know whether a given price x is high or low at a particular moment in time, we need to look back in time and observe how much transaction issuers are paying for their transactions.

\begin{figure}[H]
  \centering
  \includegraphics[width=\linewidth]{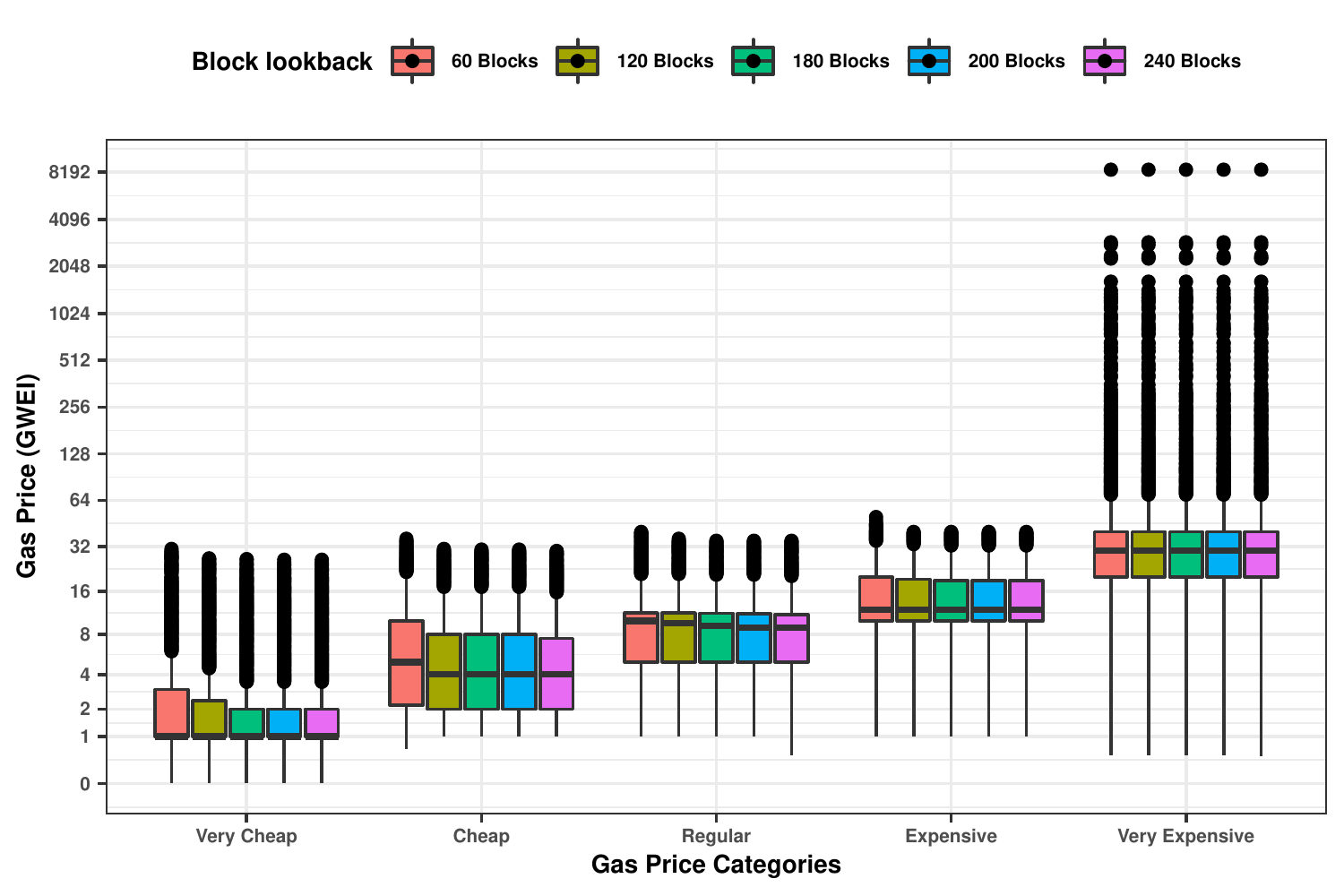}
  \caption{The effect of the block lookback parameter on the definition (distribution) of gas price categories.
  .}
  \label{fig:appendix-rq1-lookback}
\end{figure}

There is no consensus on how further back one should go to determine what the current payment norm is. EthGasStation, the most popular gas tracker, provides gas price statistics based on the past 200 blocks\footnote{\url{https://ethgasstation.info/txPoolReport.php}}. Looking back 200 blocks is equivalent to looking back 200 * 15 seconds = 3,000 seconds = 50 minutes. In this study, we look back 120 blocks. Our rationale was to cover 30 minutes (half an hour). We believe that, in practice, looking back half an hour is more intuitive and straightforward than looking back 50 minutes.
We performed a sensitivity analysis in order to understand how the block lookback choice influences the definition (distribution) of our five gas price categories. We evaluate the following block lookback choices: 60, 120, 180, 200 (EthGasStation) and 240. The results are shown in Figure~\ref{fig:appendix-rq1-lookback}.

As depicted in Figure~\ref{fig:appendix-rq1-lookback}, the definition of the gas price categories is only subtly affected by the block lookback choice. Therefore, we believe that our conclusions are likely to hold to all the five lookback choices investigated in this sensitivity analysis. Further studies in the topic should more extensively evaluate the impact of the choice of the block lookback on the definition of gas prices and on the prediction of transaction processing times.
\newpage

\section{RQ2: Summary of accuracy statistics for the prediction models}
\label{appendix:rq2}

\begin{table}[H]
  \centering
  \caption{Summary of accuracy statistics for the prediction models (MAE = Mean Absolute Error, MedAE = Median Absolute Error, MAPE = Mean Absolute Percentage Error, and MedAPE = Median Absolute Percentage Error). Values shown in minutes.}
  \label{tab:summary-models-rq2}
  \begin{tabular}{lrrrr}
  \hline
  \textbf{Model} & \multicolumn{1}{l}{\textbf{MAE}} & \multicolumn{1}{l}{\textbf{MedAE}} & \multicolumn{1}{l}{\textbf{MAPE}} & \multicolumn{1}{l}{\textbf{MedAPE}} \\ \hline
  Etherscan Pending Tx. Page    & 70.10 & 0.97 & 3062.24 & 74.64 \\
  Etherscan Gas Tracker Page    & 42.12 & 0.82 & 770.34  & 74.27 \\
  EthGasStation Gas Price API   & 67.07 & 0.68 & 2695.43 & 62.18 \\
  EthGasStation Pred. Table API & 64.28 & 0.83 & 2854.67 & 64.37 \\ \hline
  \end{tabular}
\end{table}

\begin{table}[H]
  \centering
  \caption{Summary of accuracy statistics for the prediction models -- per gas price category (MAE = Mean Absolute Error, MedAE =Median Absolute Error, MAPE = Mean Absolute Percentage Error, and MedAPE = Median Absolute Percentage Error). Values shown in minutes.}
  \label{tab:summary-models-rq2-per-price}
  \begin{tabular}{llrrrr}
  \hline
  \textbf{Model} &
    \textbf{\begin{tabular}[c]{@{}l@{}}Gas Price \\ Category\end{tabular}} &
    \multicolumn{1}{l}{\textbf{MAE}} &
    \multicolumn{1}{l}{\textbf{MedAE}} &
    \multicolumn{1}{l}{\textbf{MAPE}} &
    \multicolumn{1}{l}{\textbf{MedAPE}} \\ \hline
  Etherscan Pending Tx. Page    & \multirow{4}{*}{Very Cheap}     & 182.39 & 11.78  & 5649.20 & 239.60  \\
  Etherscan Gas Tracker         &                                 & 117.47 & 7.12   & 2033.64 & 97.86   \\
  EthGasStation Gas Price API   &                                 & 197.18 & 209.23 & 7790.10 & 1316.70 \\
  EthGasStation Pred. Table API &                                 & 159.35 & 18.91  & 5005.57 & 394.18  \\ \hline
  Etherscan Pending Tx. Page    & \multirow{4}{*}{Cheap}          & 58.05  & 2.23   & 3808.38 & 113.66  \\
  Etherscan Gas Tracker         &                                 & 25.39  & 1.60   & 822.75  & 83.40   \\
  EthGasStation Gas Price API   &                                 & 50.26  & 2.89   & 3232.69 & 88.20   \\
  EthGasStation Pred. Table API &                                 & 61.61  & 2.61   & 4169.47 & 130.44  \\ \hline
  Etherscan Pending Tx. Page    & \multirow{4}{*}{Regular}        & 38.76  & 0.83   & 2047.16 & 73.94   \\
  Etherscan Gas Tracker         &                                 & 24.02  & 0.67   & 335.40  & 71.36   \\
  EthGasStation Gas Price API   &                                 & 26.30  & 0.47   & 614.76  & 54.50   \\
  EthGasStation Pred. Table API &                                 & 40.74  & 0.62   & 2560.22 & 61.26   \\ \hline
  Etherscan Pending Tx. Page    & \multirow{4}{*}{Expensive}      & 24.09  & 0.37   & 1786.01 & 50.18   \\
  Etherscan Gas Tracker         &                                 & 15.22  & 0.42   & 159.01  & 65.94   \\
  EthGasStation Gas Price API   &                                 & 14.86  & 0.23   & 89.64   & 41.58   \\
  EthGasStation Pred. Table API &                                 & 23.36  & 0.27   & 1497.69 & 43.28   \\ \hline
  Etherscan Pending Tx. Page    & \multirow{4}{*}{Very Expensive} & 7.37   & 0.25   & 1175.75 & 38.37   \\
  Etherscan Gas Tracker         &                                 & 1.30   & 0.38   & 65.24   & 66.10   \\
  EthGasStation Gas Price API   &                                 & 1.11   & 0.20   & 40.01   & 35.21   \\
  EthGasStation Pred. Table API &                                 & 2.47   & 0.20   & 278.85  & 36.50   \\ \hline
  \end{tabular}
\end{table}

\newpage

\section{Post-hoc study: Summary of accuracy statistics for the prediction models}
\label{appendix:post-hoc-study}


\begin{table}[H]
    \centering
    \caption{Summary of accuracy statistics for the prediction models (MAE = Mean Absolute Error, MedAE = Median Absolute Error, MAPE = Mean Absolute Percentage Error, and MedAPE = Median Absolute Percentage Error). Values shown in minutes.}
    \label{tab:summary-models-posthoc}
    \begin{tabular}{lrrrr}
    \hline
    \textbf{Model} & \multicolumn{1}{l}{\textbf{MAE}} & \multicolumn{1}{l}{\textbf{MedAE}} & \multicolumn{1}{l}{\textbf{MAPE}} & \multicolumn{1}{l}{\textbf{MedAPE}} \\ \hline
    Our Linear Regression Model & 2.01 & 0.53 & 78.31 & 49.98 \\
    The State-of-the-Practice Model & 14.78 & 0.70 & 861.88 & 59.44 \\ \hline
    \end{tabular}
  \end{table}


\begin{table}[H]
    \centering
    \caption{Summary of accuracy statistics for the prediction models -- per gas price category. (MAE = Mean Absolute Error, MedAE = Median Absolute Error, MAPE = Mean Absolute Percentage Error, and MedAPE = Median Absolute Percentage Error). Values shown in minutes.}
    \label{tab:summary-models-posthoc-per-price}
    \begin{tabular}{llrrrr}
    \hline
    \textbf{Model} &
      \textbf{\begin{tabular}[c]{@{}l@{}}Gas Price \\ Category\end{tabular}} &
      \multicolumn{1}{l}{\textbf{MAE}} &
      \multicolumn{1}{l}{\textbf{MedAE}} &
      \multicolumn{1}{l}{\textbf{MAPE}} &
      \multicolumn{1}{l}{\textbf{MedAPE}} \\ \hline
    Our Linear Regression Model    & \multirow{2}{*}{Very Cheap}     & 5.63 & 2.35  & 118.02 & 69.22  \\
    The State-of-the-Practice Model &                                 & 44.65 & 6.73  & 2164.73 & 97.58  \\ \hline
    Our Linear Regression Model    & \multirow{2}{*}{Cheap}          & 1.90  & 0.93   & 86.19 & 57.35  \\
    The State-of-the-Practice Model &                                 & 13.29  & 2.02   & 900.85 & 84.24  \\ \hline
    Our Linear Regression Model    & \multirow{2}{*}{Regular}        & 0.92  & 0.49   & 76.24 & 51.14   \\
    The State-of-the-Practice Model &                                 & 7.40  & 0.59   & 751.61 & 57.65   \\ \hline
    Our Linear Regression Model    & \multirow{2}{*}{Expensive}      & 0.47  & 0.29   & 58.68 & 40.62   \\
    The State-of-the-Practice Model &                                 & 1.10  & 0.25   & 112.13 & 42.40   \\ \hline
    Our Linear Regression Model    & \multirow{2}{*}{Very Expensive} & 0.29   & 0.20   & 42.96 & 35.20   \\
    The State-of-the-Practice Model &                                 & 0.30   & 0.19   & 41.54  & 36.00   \\ \hline
    \end{tabular}
\end{table}


\newpage

\end{document}